\documentclass{jfm}
\usepackage{graphicx}
\usepackage{epstopdf, epsfig}
\usepackage{amsmath}
\usepackage{caption}
\usepackage{subcaption}
\usepackage{todonotes}
\usepackage{bm}

\usepackage[normalem]{ulem}
\usepackage{color}

\shorttitle{Instability of the optimal edge trajectory in the Blasius boundary layer}
\shortauthor{M. Beneitez, Y. Duguet, P. Schlatter and D. S. Henningson}

\title{Instability of the optimal edge trajectory in the Blasius boundary layer}

\author{Miguel Beneitez\footnote{Current address: DAMTP, Centre for Mathematical Sciences, Wilberforce Road, Cambridge CB3 0WA, UK}\aff{1}
  \corresp{\email{beneitez@kth.se}},
  Yohann Duguet\aff{2},
  Philipp Schlatter\aff{1}
 \and Dan S. Henningson\aff{1}}

\affiliation{ \aff{1}FLOW Centre and Swedish e-Science Research Centre (SeRC), KTH Engineering Mechanics, Royal Institute of Technology, SE-100 44 Stockholm, Sweden
\aff{2}LISN-CNRS, Campus Universitaire d'Orsay, Universit\'e Paris-Saclay, F-91400, Orsay, France}

\begin{document}

\maketitle

\begin{abstract}

In the context of linear stability analysis, considering unsteady
base flows is notoriously difficult. A  generalisation of modal linear
stability analysis, allowing for arbitrarily unsteady base flows over
a finite time, is therefore required. The recently developed optimally
time-dependent (OTD) modes form a projection basis for the tangent
space. They capture the leading amplification directions in state
space under the constraint that they form an orthonormal basis at all
times. The present numerical study illustrates the possibility to describe a complex flow case using the leading OTD modes. The flow under investigation is an unsteady case of the Blasius boundary layer, featuring streamwise streaks of finite length and relevant to bypass transition. It corresponds to the state space trajectory initiated by the minimal
seed; such a trajectory is unsteady, free from any spatial
symmetry, and shadows the laminar-turbulent separatrix for a finite
time only. The finite-time instability of this unsteady base flow is investigated
using the 8 leading OTD modes. The analysis includes the computation of finite-time Lyapunov
exponents as well as instantaneous eigenvalues, and of the associated flow structures. The reconstructed instantaneous eigenmodes are all of outer type. They
map unambiguously the spatial regions of largest instantaneous
growth. Other flow structures, previously reported as secondary, are
identified with this method as relevant to streak switching and to
streamwise vortical ejections. The dynamics inside the
tangent space features both modal and non-modal amplification. Non-normality within the
reduced tangent subspace, quantified by the instantaneous numerical
abscissa, emerges only as the unsteadiness of the base flow is reduced.

\end{abstract}


\section{Introduction}

Hydrodynamic stability theory aims at characterising the stability of a given base flow
to infinitesimal or finite-amplitude disturbances. In most academic cases, the base flow of interest is known analytically and is generally independent of time \citep{drazin2004hydrodynamic}. There are however physical contexts in which 
the choice of a physically relevant base flow is not obvious. Bypass transition to turbulence in shear flows
falls into this category: there is ample experimental and numerical evidence that turbulent fluctuations emerge from the breakdown of laminar streamwise streaks
of sufficiently strong amplitude \citep{morkovin1969many,henningson1993mechanism,matsubara2001disturbance,jacobs2001simulations,brandt2002transition,brandt2004transition} rather than from the destabilisation of the steady laminar base flow. Streamwise streaks, originally called Klebanoff modes, are loosely defined as spanwise modulations of the streamwise velocity field \citep{klebanoff1962three}. They are predominantly streamwise-independent structures supporting three-dimensional wiggles convected at different velocities \citep{brandt2003convectively}.  Streaks are not associated mathematically to unstable eigenmodes of the purely laminar base flow, instead they emerge because of the non-normality of the associated linear operator \citep{schmid2001stability} via a mechanism called lift-up. This mechanism transfers streamwise vorticity upstream into streaks further downstream \citep{landahl1980note,brandt2014lift}. Careful early experiments have suggested that their breakdown follows an instability mechanism \citep{bakchinov1995transition,alfredsson1996streaky,matsubara2001disturbance,asai2002instability}. The exact temporal dynamics of finite-amplitude streaks is however not trivial. In several numerical studies, a frozen  (two-dimensional) finite-amplitude streak pattern was considered as a base flow, and its linear stability analysis was carried out by assuming that the perturbations are inviscid \citep{andersson2001breakdown,kawahara2003linear,brandt2007numerical}. The unstable eigenfunctions identified break the translational invariance of the initial streaks. The main outcome of the stability analysis of \emph{streamwise-invariant streaks} is the possibility for two different ways of breaking this streamwise invariance, either by symmetric (varicose) or anti-symmetric (sinuous) eigenmodes. Around that time, \cite{hamilton1995regeneration} made use of the concept of subcritical streak instability to justify the three-dimensionality of the self-sustaining process in all shear flows \citep{hamilton1995regeneration,waleffe1997self}. 
\cite{hoepffner2005transient}, following \cite{schoppa2002coherent},
showed that streamwise modulations of the streaks observed during transition, although possible as a linear instability of the frozen streaks, can also arise for lower streak amplitudes via non-normal amplification of streak disturbances over a finite-time. In a related study, the secondary instability of time-dependent streaks in channel flow was addressed by adopting a finite-time formalism by \cite{cossu2007optimal}. \cite{schlatter2008streak} studied the secondary instability of streaks via nonlinear impulse response. 
Linear stability features were later extracted directly from numerical data \citep{vaughan2011stability,hack2014streak} by considering an instantaneous streamwise-independent base flow. More recently, the stability of streaks in turbulent flows was also considered by focusing on the associated mean flow rather than on instantaneous flow fields \citep{alizard2015linear,cassinelli2017streak}. It remains hence an open question whether there are additional insights for stability analysis by considering fully unsteady three-dimensional base flows. This paper is devoted to a computational exploration of the possibilities offered by this approach.

In the context of initial value problems, an initial condition at time $t=t_0$ is represented by a point in the associated state space. The knowledge of a given initial condition defines uniquely the \emph{base flow}, i.e. the unsteady state space trajectory initiated by that particular initial condition. In principle, the arbitrary unsteadiness of the base flow is not an obstacle to modal linear stability analysis (LSA), at least when the base flow corresponds to an attractor defined over unbounded times. The generalisation of eigenvalues is given by (time-independent) Lyapunov exponents (LEs), defined as ergodic averages of the instantaneous divergence rate between trajectories \cite{viana2014lectures}. The generalisation of the eigenvectors is given by the (time-dependent) covariant Lyapunov vectors (CLVs) \cite{ginelli2007characterizing,kuptsov2012theory,pikovsky2016lyapunov}. Eventually, in the present study, an additional theoretical limitation is the requirement that the method be applicable to a base flow defined only over a finite-time interval. This requirement is made necessary by the convective nature of the 
boundary layer and the fact that any spatially localised perturbation to the Blasius flow has to exit a bounded computational domain in a finite time. In this context, most infinite-time concepts such as eigenvalues need to be formally redefined over the finite time interval of interest. While this does not pose any strong mathematical difficulty, it crucially determines the mathematical toolbox relevant for that problem.

We are interested here in a base flow featuring streamwise streaks of finite length and width, with an unsteady dynamics. Since we wish to define the base flow in an unambiguous way, it is initialised at $t=0$ from a well-defined finite-amplitude perturbation to the original laminar Blasius flow. In the present context of identifying the mechanisms allowing for transition from a minimal level of disturbance, the selected initial condition is the laminar base flow, perturbed at $t=0$ by the so-called minimal seed \cite{Kerswell:2018aa}. The minimal seed is defined rigorously as the disturbance of lowest energy capable of triggering turbulence, or equivalently the point on the edge manifold closest to the laminar attractor in energy norm \cite{Kerswell:2018aa}. Its computation is based on a nonlinear optimization method \citep{pringle2010using,Cherubini_2011,vavaliaris2020optimal} and in practice requires an optimization time interval $(0,T_{\text{opt}})$. The trajectory initiated by this flow field is called optimal edge trajectory. By construction it is an edge trajectory i.e. it belongs to the invariant set called the laminar-turbulent boundary: some infinitesimal perturbations to such trajectories lead to relaminarisation while others trigger turbulent flow. The concept of edge trajectory was originally introduced in bistable parallel shear flows \citep{itano2001dynamics}: the asymptotic fate of such edge trajectories form the edge state, a relative attractor in state space, whose stable manifold divides the state space in two disjoint and complementary basins. Its extension to boundary layer flows is trivial for parallel boundary layer flows \citep{khapko2013localized,khapko2014complexity,khapko2016edge,biau2012laminar} but less straightforward in spatially developing boundary layer flows like the Blasius boundary layer \citep{Cherubini_2011,Duguet2012}. In such cases the concept of a turbulent attractor is not clearly defined, yet edge trajectories can still be identified, at least over finite times. In boundary layers, the edge concept becomes fragile on very long timescales because the laminar Blasius flow can develop instabilities to Tollmien--Schlichting waves over long time horizons \cite{beneitez2019edge,beneitez2020modeling}. In the absence of an asymptotic state, the stability of finite-time edge trajectories cannot be investigated using Lyapunov exponents and CLVs, all based on ergodic infinite-time averages. The generalisations of eigenvalues/LEs on finite times are, trivially, the finite-time Lyapunov exponents (FTLEs). Their large-time limits, when they are defined, coincide indeed with LEs \cite{Haller_2015}. The eigenvectors do not, however, admit any simple finite-time generalisation.  

We chose for this task the optimally-time dependent (OTD) modes introduced recently by \cite{babaee2016minimization}. The associated formalism has two advantages: it computes physically meaningful directions in the tangent space, and yields accurate numerical estimates of the FTLEs. OTD modes approximate the linearised dynamics \cite{blanchard2019analytical} around the base flow trajectory in an optimal way, yet under the constraint that the modes remain orthogonal at all times. Orthogonality is not a property shared by CLVs. Handling an orthogonal basis is in practice a strong technical advantage over ill-conditioned bases. The trade-off is that the OTD modes do \emph{not} fulfill the covariance property. Note that, when both are defined, the leading OTD mode still coincides with the leading CLV for sufficiently long times. The reduced linearised operator, obtained by projecting the original operator on the $r$ first OTD modes, can be used to estimate the stability characteristics of the high-dimensional problem, otherwise prohibitively expensive to compute.
In particular, the eigenvalues of this reduced-order operator yield an accurate approximation of the FTLEs of the full system \citep{babaee2017}. Besides, whereas the OTD modes themselves are not interpretable physically, instantaneous eigenmodes can be reconstructed in physical space from the diagonalisation of the reduced order operator.
As shown by \cite{babaee2016minimization} from specific examples, over shorter time horizons well-initialised OTD modes can capture the non-normality of the underlying dynamics.
These properties make OTD modes an interesting tool specifically for transient phenomena. On a technical level, their implementation requires neither solutions of the adjoint system, nor data to be input, and no iterative scheme: the OTD modes are computed in \emph{real time} together with the time-evolving base flow. They however need to be initialised at $t=0$. There is currently no accepted general way of choosing initial conditions for these modes, although it is expected that past some finite transient time the OTD directions naturally align with the most important directions of the system. OTD modes have been used recently in several hydrodynamic applications, including the identification of bursting phenomena \citep{farazmand2016dynamical}, the control of linear instabilities \citep{blanchard2019control} and the stability of pulsating Poiseuille flow \citep{kern2021} as well as for faster edge tracking in high dimension \citep{beneitez2020lcs}. The current investigation, motivated by these promising properties, is an opportunity to test a new computational framework for stability calculations considered until now as challenging.

The present study revisits the optimal edge trajectory in the Blasius boundary layer by considering it as the new finite-time base flow, and by determining its stability characteristics using the new finite-time framework offered by OTD modes. In particular, the physical structure of the leading modes will be analysed at different times, with a focus on the influence of the time dependence of the base flow on the results. The structure of this paper is as follows. The OTD modes are introduced mathematically in a general context in Section 2. The computational set-up, the implementation, and the details of the reference edge trajectory are described in Section 3. Section 4 contains the stability analysis using the proposed methodology. Finally, the conclusions are given and discussed in Section 5.

\section{Theoretical framework \label{sec:otd}} 

\subsection{Linearisation around an arbitrary base flow \label{sec:bf}}

The context of the current study is very general. Assuming that a spatially discretised flow field can be represented by $n$ independent real-valued degrees of freedom with $n\gg 1$ (see e.g. \citep{gibson_halcrow_cvitanovic_2008}), we consider $\mathbb{R}^{n}$ as the original high-dimensional space of reference.
We suppose a non-autonomous dynamical system defined over a time interval $[t_0,t_1)$:
\begin{equation}
    \frac{d{\bm Q}}{dt}={\bm f}({\bm Q},t),\quad 
    {\bm Q}(t_0)={\bm Q}_0,
     \label{eq:main}
\end{equation}
where ${\bm Q}_0,{\bm Q}\in \mathbb{R}^{n}$, and ${\bm f}:\mathbb{R}^{n} \rightarrow \mathbb{R}^{n}$ is a diffeomorphism. We suppose both $t_0$ and $t_1$ finite although $t_1\to+\infty$ is also possible.
For a given choice of ${\bm Q}_0$, we define the solution to Eq. \eqref{eq:main}, namely $\bar{{\bm Q}}:(t_0:t_1)\rightarrow \mathbb{R}^{n}$ as the \emph{base flow} whose stability we will now determine.

Let ${\bm q}(t)$ represent a small perturbation to $\bar{{\bm Q}}(t)$, small enough so that the dynamics can be linearised around $\bar{{\bm Q}}(t)$ (mathematically ${\bm q}$ evolves in the \emph{tangent space} associated with the dynamics). Then ${\bm q}$ is governed by the linearised equation
\begin{equation}
    \frac{d{\bm q}}{dt}={\bm L}(\bar{{\bm Q}},t){\bm q},\ \quad {\bm L}(\bar{{\bm Q}},t)={\bm \nabla}_{\bm Q} {\bm f}(\bar{{\bm Q}},t),
    \label{dqdt}
\end{equation}
where ${\bm \nabla}_{\bm Q} {\bm f}(\bar{{\bm Q}},t)$ is the $n \times n$ (time-dependent) Jacobian matrix, evaluated along the base flow at time $t$. The OTD modes, to be introduced in Section \ref{sec:otdmethod}, form a basis of time-dependent real-valued vectors (a complex-valued definition is also possible, but is not discussed here). They approximate in an optimal way the leading directions of the Jacobian operator. They are better understood after the notion of covariant vectors is discussed in Section \ref{sec:cov}.

\subsection{Covariance property \label{sec:cov}}

An ideal basis for the linearised dynamics should allow one to split the whole $n$-dimensional tangent space into a direct sum of subspaces evolving along the flow, each one with its own specific dynamics \citep{pikovsky2016lyapunov}. The associated time-dependent directions spanning these subspaces are referred to as dynamically \emph{covariant}. If the base flow $\bar{Q}$ does \emph{not} depend on time, the covariance property classically defines expanding and contracting eigenspaces, the covariant vectors are the associated eigenvectors, and that the temporal rate-of-change of their norm defines the eigenvalues. In the general case, these vectors are called covariant Lyapunov vectors (CLVs) or sometimes simply Lyapunov vectors. By definition, CLVs can be re-interpreted as zeros of the functional 
\begin{equation}
    \mathcal{J}=\lim_{\delta t\to 0}\frac{1}{(\delta t)^2}\sum_{i=1}^{n}||{\bm w}_i(t+\delta t)-({\bm \nabla} {\bm F}_{t}^{t+\delta t}){\bm w}_i(t)||^2, \label{eq:functional}
\end{equation}
where ${\bm F}_{t}^{t+\delta t}:\mathbb{R}^{n} \rightarrow \mathbb{R}^{n}$ is the infinitesimal forward propagator associated with Eq. \eqref{eq:main}. The Jacobian matrix ${\bm \nabla} {\bm F}_{t}^{t+\delta t}$ maps a vector of the tangent space $\bm{w}_i(t)$ at time $t$ to its image at the later time $t+\delta t$ in the corresponding tangent space. 

The issues associated with CLVs are two-fold. First, they are not necessarily mutually orthogonal at a given time, making the associated basis possibly ill-conditioned. Second, the only algorithms known to compute them are proven to be valid only on attractors on which the dynamics is ergodic \citep{ginelli2007characterizing,kuptsov2012theory}. CLVs are thus essentially an inappropriate computational tool for the study of transients. 

\subsection{Optimally time-dependent modes \label{sec:otdmethod}}

\begin{figure}
    \centering
    \includegraphics[width=0.85\textwidth]{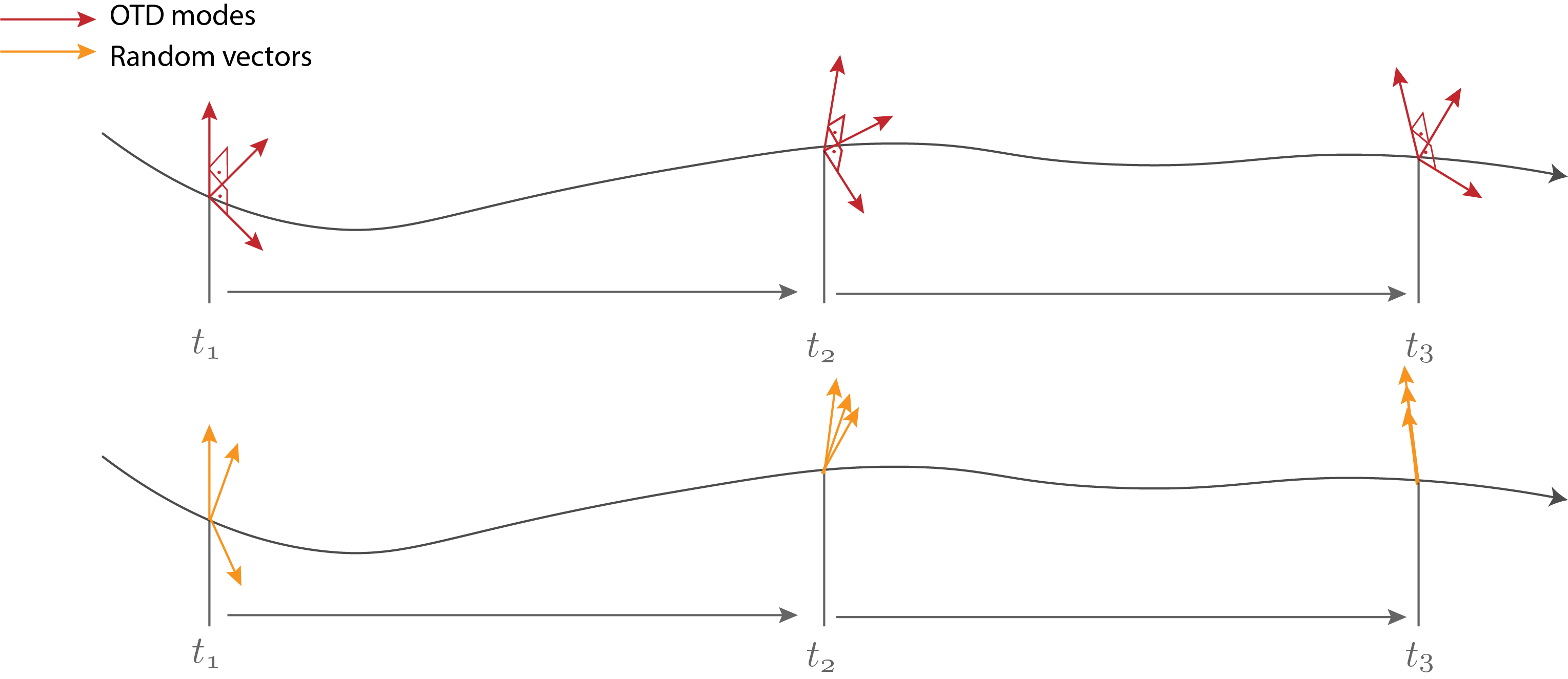}
    \caption{Sketch of different vector bases for the tangent space of a given unsteady base flow. Temporal evolution of respectively OTD modes (red, top) and random unit vectors (orange, bottom) propagated along a base flow trajectory. The leading non-rescaled direction coincides with the leading OTD mode and is shown pointing upwards at all times.}
    \label{fig:clv_comps}
\end{figure}

\cite{babaee2016minimization} defined the OTD modes ${\bm u}_i,~i=1,...,r$ as a computational compromise. These are minimisers of the functional $\mathcal{J}$ in Eq. \eqref{eq:functional}, under the additional constraint that they form an orthonormal basis at all times. The orthonormality constraint simply reads
\begin{equation}
     \langle {\bm u}_i(t), {\bm u}_j(t)\rangle=\delta_{ij},~i,j=1,..,r, \qquad r\ll n,
     \label{eq:ortho}
\end{equation}
where $\langle \cdot,\cdot \rangle$ denotes the inner product associated with the $L^2$ norm and $\delta_{ij}$ is the classical Kronecker symbol. The time evolutions of the OTD modes and of a set of initially random unit vectors are depicted schematically in Fig.~\ref{fig:clv_comps} for illustration purposes. Random unit vectors would follow the tangent dynamics and all align rapidly with the most expanding direction, making them poor candidates to describe and analyse the dynamics of the tangent space. OTD modes follow the tangent dynamics but stay orthonormal at all times, avoiding any alignment issues which might occur using e.g. CLVs. This constraint destroys the interpretability of each OTD direction in terms of covariant dynamics. However, the orthonormality of the OTD modes is particularly appealing for reduced order modelling, for instance in the context of control \citep{blanchard2019analytical,blanchard2019control}.

As derived in \cite{babaee2016minimization}, the maximisation of $\mathcal{J}$ in Eq. \eqref{eq:functional} under the orthogonality constraint (\ref{eq:ortho}) yields a system of coupled nonlinear evolution equations:
\begin{equation}
    \frac{{d\bm u}_i}{dt} = {\bm L}(t) {\bm u}_i-\sum_{j=1}^r [\langle {\bm L}(t) {\bm u}_i, {\bm u}_j\rangle-{\bm A}_{ij}(t)]{\bm u}_j,~i=1,..,r \label{eq:otd}
\end{equation}
The nonlinearity is a direct consequence from the orthonormality constraint. The matrix $\mathbf{A}\in \mathbb{R}^{r\times r}$ refers \emph{a priori} to any skew-symmetric matrix. The discretised equations eq. \eqref{eq:otd} for $i=1,...,r$ form, together with eq. \eqref{eq:main}, a closed $(r+1)$-dimensional system of real-valued ODEs. The ${\bm u}_i$'s depend on the instantaneous vector $\bar{\mathbf{Q}}(t)$, however $\bar{\mathbf{Q}}$ itself is unaffected by the evolution of the ${\bm u}_i$'s. In \cite{blanchard2019analytical}, a particular choice of $\mathbf{A}$ was made:
\[
{\bm A}_{ij} = \Bigg\{ \begin{array}{c}
     -\langle \bm {L} {\bm u}_j, {\bm u}_i \rangle, \ j<i\\
    0,\ i=j \\
    \langle \bm L {\bm u}_i, {\bm u}_j \rangle, \ j>i,
\end{array}
\]
which was also considered here. An arbitrary choice of $\mathbf{A}$ would lead to a fully coupled system so that each ${\bm u}_i$ appears in every equation in eq. \eqref{eq:otd}. However, under the current choice the set of $r$ equations in eq. \eqref{eq:otd} has a lower triangular form: the evolution of the $i^{th}$ mode depends only on the modes from $1$ to $i$, making the OTD formulation hierarchical. The resulting evolution equation for each OTD mode becomes 
\begin{equation}
    \frac{d{\bm u}_i}{dt} = {\bm L}(t) {\bm u}_i-\langle {\bm L}(t) {\bm u}_i, {\bm u}_i \rangle {\bm u}_i-\sum_{j=1}^{i-1} [\langle {\bm L}(t) {\bm u}_i, {\bm u}_j\rangle + \langle {\bm L}(t) {\bm u}_j, {\bm u}_i\rangle ]{\bm u}_j. \label{eq:otdblanchard} 
\end{equation}

The nonlinear system of $r$ equations (\ref{eq:otdblanchard}) can be evolved forward in time together with Eq. (\ref{eq:main}), from which the matrix ${\bm L}$ can be evaluated at all times. Eq. (\ref{eq:main}) remains however independent of the evolution of each ${\bm u}_i$. This results in an $(r+1)\times n$-dimensional asymmetrically coupled dynamical system. The OTD modes retain a (short-time) memory of their initial conditions. They are in general not covariant, except for base flows such that all instantaneous eigenvectors remain normal to each other.

\subsection{The reduced linearised operator  \label{sec:Lr}}

In order to analyse the linearised dynamics {\it within} the reduced subspace optimally spanned by the OTD modes, \cite{babaee2016minimization} introduced the reduced operator ${\bm L}_r$ defined by projecting the high-dimensional operator ${\bm L}$ onto the OTD directions:
\begin{equation}
     \bm L_{r_{ij}}(t) = \langle {\bm u}_i,{\bm L}(t){\bm u}_j\rangle\quad i,j=1,\dots,r.\\ \label{eq:reduced_operator}
\end{equation}
In particular all the instantaneous stability indicators defined in the next subsection will be derived from algebraic properties of the $r \times r$ matrix ${\bm L}_r$ evaluated at the relevant times.

For a time-independent linearised operator $L$ the space spanned by the modes $\left\{u_i\right\}_{i=1}^{r}$ converges asymptotically to the most unstable eigenspace of $L$. Moreover, if $L$ happens to be also symmetric, the OTD modes coincide  with its eigenvectors at all times. 

\subsection{Instantaneous stability indicators \label{sec:eigenmodes}}

As emphasized in \cite{babaee2016minimization}, although they correspond to divergence-free vector fields the ${\bm u}_i$'s do not have a direct physical interpretation as flow fields. More meaningful vector sets can nevertheless be reconstructed instantaneously from the knowledge of the ${\bm u}_i$'s and the reduced operator. At every time, $L_r(t)$ can be diagonalised as ${\bm L}_r={\bm E}^{\lambda}{\bm \Lambda}_r({\bm E}^{\lambda})^{-1}$ with ${\bm E}^{\lambda}$ and ${\bm \Lambda}_r=\text{diag}(\lambda_1(t),...,\lambda_r(t))$. The new modes ${\bm u}_i^{\lambda}, i=1,...,r$ are defined (using the summation convention) by 
\begin{equation}
    \label{eq:Ulambda}
    {\bm u}_i^{\lambda}(t)=\bm E^{\lambda}_{ij}(t){\bm u}_j(t). 
\end{equation}

Unlike the ${\bm u}_i$'s, the ${\bm u}_i^{\lambda}$'s are not necessarily mutually orthogonal. They are interpreted as \emph{instantaneous eigenmodes}. They are the only velocity fields used for visualisation in this paper. We emphasize that, although the ${\bm u}_i$'s are real-valued, the modes ${\bm u}_i^{\lambda}$ are complex-valued and come in pairs. Only the real parts, with arbitrary phase, of the associated velocity fields will be represented. The time-dependent numbers $\lambda_1(t),...,\lambda_r(t),~i=1,...,r$ are labelled \emph{instantaneous eigenvalues} and are complex-valued.

Another key scalar quantity is the instantaneous numerical abscissa $\sigma(t)$, a positive number,  defined as the largest eigenvalue of the symmetrised reduced operator $({\bm L}+{\bm L}^T)/2$ \citep{embree2005spectra}. $\sigma$ corresponds to the largest possible growth rate at a given time, due to both normal and non-normal effects combined together. Whenever ${\bm L}$ is non-normal, $\sigma$ is strictly larger than the real part of all $\lambda_i$'s. 

For a given integer value $r$, it is a natural extension to define $\sigma_r$ as the largest eigenvalue of the symmetrised reduced operator $({\bm L}_r+{\bm L}_r^T)/2$. The gap
\begin{eqnarray}
g_r(t):=\text{min}_i|\sigma_r-\operatorname{Re}(\lambda_i)|=\sigma_r-\operatorname{Re}(\lambda_1)
\label{eq:gap}
\end{eqnarray}
quantifies the non-normality of the reduced linearised operator at each instant. The values of $g_r(t)$ bound from below the value of $g_{n}(t)$ corresponding to the full high-dimensional system. In the remainder of the paper we will not make a difference between $g_r$ and $g_{n}$ and will simply use the notation $g(t)$. Note that for arbitrary time-dependent operators and finite $r$, it is possible to have $\sigma_r \approx \max \lambda_i$ even for a non-normal operator.

\subsection{Finite-time Lyapunov exponents \label{sec:FTLEs}}

For a general dynamical system in dimension $n$, characterised by a propagator ${\bm F}_{t_0}^{t}$, the Cauchy-Green tensor $\mathbf{C}_{t_0}^{t}$ is defined as 
\begin{equation}
    {\bm C}_{t_0}^{t}({\bm q}_0) :=[{\bm \nabla}  {\bm F}_{t_0}^{t}({\bm q}_0)]^T [{\bm \nabla} {\bm F}_{t_0}^{t}({\bm q}_0)].
    \label{eq:CG}
\end{equation}

The associated finite-time Lyapunov exponents (FTLEs) are defined directly from the eigenvalues $\gamma_1>\gamma_2>...>\gamma_n$ of the Cauchy-Green tensor \citep{Haller_2015}. These eigenvalues are real and positive by virtue of the positive definiteness of the Cauchy-Green tensor. Each FTLE is defined, for an initial time $t_0$ and a horizon time $T>0$ \citep{Haller_2015}, as
\begin{equation}
    (\Lambda_{t_0}^{t_0+T})_i=\frac{1}{T}\log{\sqrt{\gamma_i}},~i=1,...,n.\\
    \label{eq:FTLE}
\end{equation}

\cite{babaee2017} provided an analytical proof that for that for any integer $r>0$, the $r$-dimensional OTD-subspace aligns exponentially fast, i.e. for increasing $T$ with the space spanned by the $r$ most dominant left vectors of the Cauchy-Green tensor. The exact rate of convergence depends on the spectrum of the problem at hand. The OTD formulation is hence a robust direct method to estimate the $r$ leading finite-time Lyapunov exponents $(\Lambda_{t_0}^{t_0+T})_i,~i=1,...,r$ of the full system at any time $t_0$ \citep{babaee2017,sapsis2018new,blanchard2019analytical}, provided the time horizon $T$ is large enough. As a consequence
the FTLEs $(\Lambda_{t_{0}}^{t_{0}+T})_i,~i=1,...,r$ are evaluated simply by averaging over time the diagonal elements of ${\bm L}_r$ \citep{blanchard2019analytical}. They are expressed as 
\begin{equation}
    (\Lambda_{t_{0}}^{t_{0}+T})_i \approx \frac{1}{T}\int_{t_{0}}^{t_{0}+T}\langle {\bm u}_i(\tau),{\bm L}_r(\tau){\bm u}_i(\tau)\rangle d\tau,~i=1,...,r,
    \label{eq:FTLE2}
\end{equation}

It can be useful to relate the OTD modes to other known vector sets from the literature beyond the CLVs. The Gram-Schmidt vectors are precisely involved in the classical algorithms used for computing finite-time Lyapunov exponents and hence LEs (see e.g. \citep{shimada1979numerical}). The OTD modes coincide with the so-called Gram-Schmidt vectors, at least in the limit where the Gram-Schmidt vectors are continuously re-orthogonalised \citep{blanchard2019analytical}.  The same modes have also been called sometimes backwards Lyapunov vectors \cite{kuptsov2012theory}. Unlike the CLVs, the OTD modes depend on the choice of the inner product except for the leading mode.

\section{Computational set-up  \label{sec:computation}}

\subsection{Direct numerical simulation  \label{sec:DNS}}

The Blasius boundary layer is the incompressible flow over a semi-infinite flat plate. It develops at the leading edge of the plate in the absence of a streamwise pressure gradient. Let $x,y,z$ denote the streamwise, wall-normal and spanwise directions, respectively. $\mathbf{v}$ is the total velocity field, $\mathbf{v}_{B}=(u_B,v_B,0)$ that of the steady Blasius solution, then $\mathbf{u}:=\mathbf{v}-\mathbf{v}_{B}=(u,v,w)$ is the perturbation velocity field. All quantities are made non-dimensional using the free-stream velocity $U_ {\infty}$ and the boundary layer thickness $\delta^{*}(x):=\int_{0}^{\infty} (1-u_B(x)/U_{\infty})dy$ of the undisturbed (steady) Blasius flow. A local Reynolds number can be defined as $Re_{\delta^{*}}(x):=U_{\infty}\delta^{*}/\nu$ with $\nu$ the kinematic viscosity of the fluid. The value of $Re_{\delta^{*}_{0}}=Re_{\delta^{*}}(x=0)$ is imposed at the upstream end ($x=0$) of the computational domain, located at a finite distance downstream of the leading edge.

The boundary conditions for the edge trajectory at the wall ($y=0$) are of no-slip and no-penetration type,
\begin{equation}
     u=v=w=0 \ ,
\end{equation}
and at the upper domain boundary ($y=L_y$) of Neumann type to allow for a natural growth of the boundary layer,
\begin{equation}
    \frac{\partial u}{\partial y}=\frac{\partial v}{\partial y}=\frac{\partial w}{\partial y}=0.
\end{equation}

A fringe region located at the downstream end of the domain damps outgoing velocity perturbations consistently with the streamwise periodic boundary conditions. The fringe is imposed as a volume force $\mathbf{F}(t,x,y,z)$ of the form
\begin{equation}
    \mathbf{F}=\gamma (x)(\mathcal{{\bm U}}(x,y,z)-\mathbf{v}(t,x,y,z)),
\end{equation}
where $\gamma(x)$ is a non-negative fringe function detailed in \citet{chevalier2007simson}. The streamwise component of $\mathcal{U}(x,y,z)$ is defined as
\begin{equation}
    \mathcal{U}_x=U(x,y,z)+[U(x+x_L,y,z)-U(x,y,z)]S\left(\frac{x-x_{\text{blend}}}{\Delta_{\text{blend}}} \right),
\end{equation}
where $S(x_{\text{blend}},\Delta_{blend})$ is a blending function connecting smoothly the outflow to the inflow, and $U(x,y,z)$ solves the boundary layer equations. The wall-normal component of $\mathcal{{\bm U}}$ is obtained via the continuity equation. In the present work the fringe length is $\Delta_{\text{blend}}=600$, $x_L$=2500 and $\gamma_{\text{max}}=0.8$. 

The present approach has been successfully applied in most works referenced in \cite{chevalier2007simson}, and in several later publications including  \cite{Duguet2012,beneitez2019edge}. 
The effect of the fringe on outgoing perturbations, allowing for the simulation of spatially developing flows in the presence of periodic boundary conditions was analysed in full mathematical detail in \cite{nordstrom1999fringe}.
\begin{figure}
    \centering
    \includegraphics[width=0.7\textwidth]{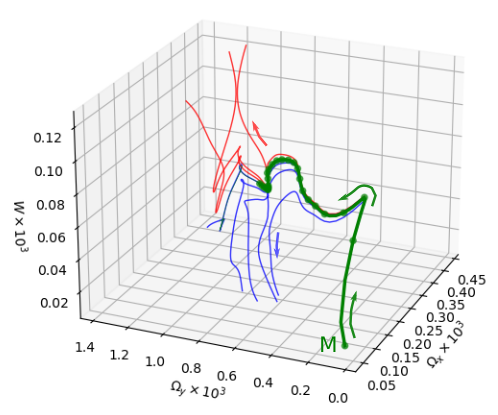}
    \caption{State portrait of the optimal edge trajectory in the Blasius boundary layer using the variables $\Omega_x$, $\Omega_y$ and $W$ defined by Eqs. \eqref{eq:observables1}--\eqref{eq:observables3}. The edge trajectory starts at $t=0$ from the minimal seed $M$ computed in \cite{vavaliaris2020optimal}. The part of the edge trajectory (green) investigated in this work, $t\in[0,800]$, is shown using a thicker line. The other trajectories start from the neighbourhood of $M$, they bracket the edge trajectory, and approach either the laminar (blue) or the turbulent state (red).  Dots are plotted every 50 time units, highlighting the slowdown already after $t=100$. $M$ indicates the location of the minimal seed.}
    \label{fig:statePortrait_ref}
\end{figure}

The temporal integration of the incompressible Navier--Stokes equations is performed using the pseudo-spectral solver SIMSON \citep{chevalier2007simson}. This direct numerical simulation (DNS) code solves the equations in the wall-normal velocity-vorticity formulation. The solution is advanced in time using a second-order Crank-Nicholson scheme for the linear terms and a fourth-order low-storage Runge-Kutta scheme for the nonlinear terms. The timestep is fixed to $\Delta t=0.2$ in terms of $U_{\infty}$ and $\delta_0^{*}$. The velocity field is expanded along $N_x$ Fourier modes in the streamwise direction $x$ and $N_z$ modes in the spanwise direction $z$, $N_y$ Chebyshev modes are used in the wall-normal direction $y$ using the Chebyshev-tau method. The evaluation of the nonlinear terms obeys the 3/2-rule for dealiasing. 

\begin{figure}
    \centering
    \includegraphics[width=0.6\textwidth]{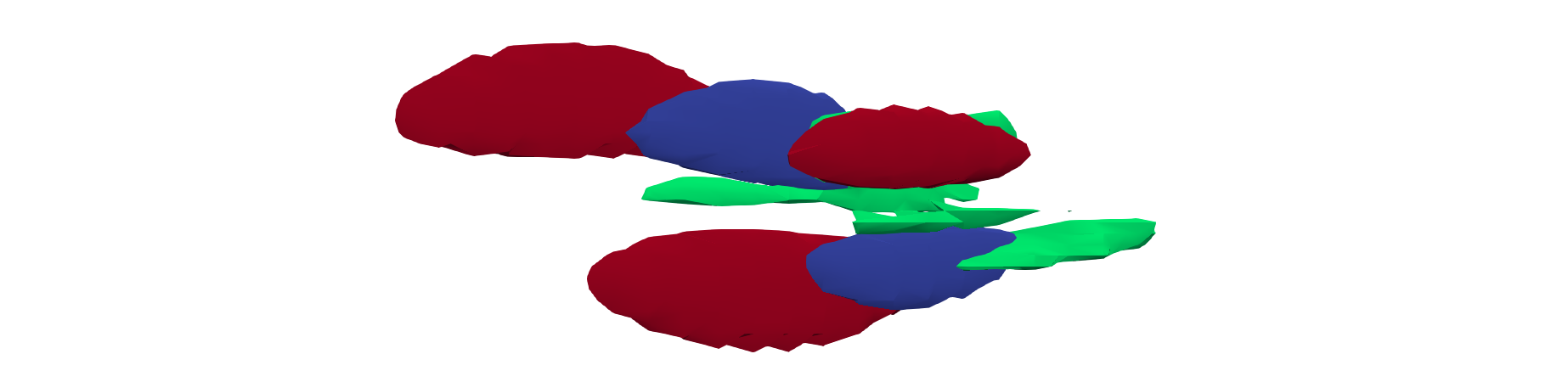}
    \caption{Three-dimensional top view of the perturbation perturbation velocity field corresponding to the initial condition of the reference trajectory. Contours of $u=-8\times 10^{-3}$ (blue), $u=6\times 10^{-3}$ (red) and global $\lambda_2^*=-4\times 10^{-4}$, where $\lambda_2^*$ denotes the vortex identification criterion introduced by \cite{jeong_hussain_1995}. The streamwise extension of the minimal seed is $\sim 22$ length units.}
    \label{fig:ICsnap}
\end{figure}

The additional equations \eqref{eq:otdblanchard} ruling the evolution of the OTD modes are advanced in time using the same scheme, based on an explicit evaluation of the inner products at every collocation point at every timestep. The initial conditions for the modes $i=1,..,r$ are spatially localised disturbance velocity fields, consistent with the localised nature of the perturbations to the streaks observed in bypass transition. The boundary conditions for equations \eqref{eq:otd} are the same as for the original DNS. The choice of boundary conditions is particularly sensitive for the perturbation equations. Further details can be found in Appendix \ref{sec:implementation}. The computational requirements for each individual OTD mode are the same of a full DNS.

Although, as described in \citet{beneitez2019edge}, it would be possible to use a moving box technique to track localised disturbances over long time horizons using limited computational resources, this is not required here because of the limited tracking time. The reference frame is hence understood as the \emph{laboratory} frame. The computational set-up for the edge tracking is similar to that in \cite{vavaliaris2020optimal}. The  computational domain $\Omega$ has dimensions $[L_x,L_y,L_z]= [2500,60,100]$ and the velocity field is expanded on $[N_x,N_y,N_z]=[2048,201,256]$ modes before dealiasing. This numerical resolution is comparable locally to that used in \cite{Duguet2012} and \cite{beneitez2019edge}. The computation of the OTD modes starts at initial time $t=$0 from the (spatially localised) minimal seed computed in \cite{vavaliaris2020optimal}. It ends at $t=800$, at which time the localised perturbation has not yet left the computational domain.

The computation of the OTD modes in \eqref{eq:otd} depends on the definition of the inner product, chosen here as
\begin{equation}
     \langle {\bm{u}},{\bm {u'}} \rangle = \int_{\Omega} (uu'+vv'+ww')d\Omega,
\end{equation}
where $\bm{u}=(u,v,w)$ and $\bm{u'}=(u',v',w')$ are any two flow fields with finite $L^2$ norm, and 
$d\Omega=dxdydz$ is the usual infinitesimal integration element over the numerical domain $\Omega$.

\subsection{The optimal edge trajectory}

\begin{figure}
    \centering
    \includegraphics[width=0.6\textwidth]{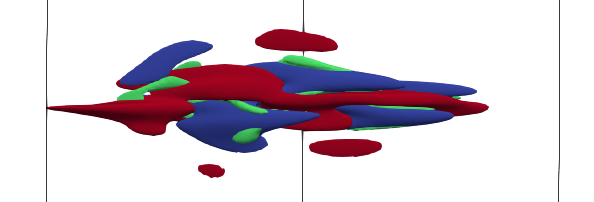}\\
    (a)\vspace{4mm}\\
    \includegraphics[width=0.9\textwidth]{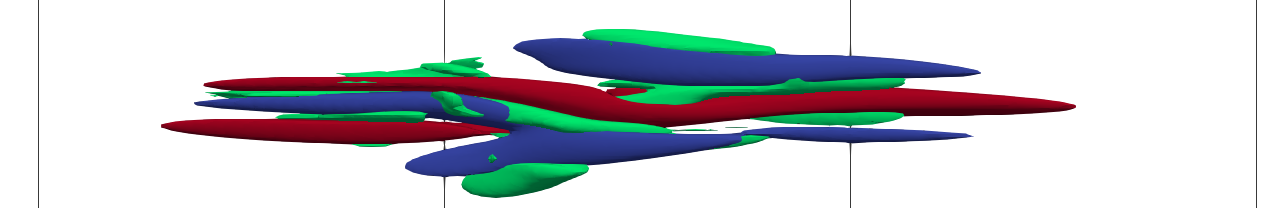}\\
    (b)\vspace{4mm}\\
    \includegraphics[width=0.9\textwidth]{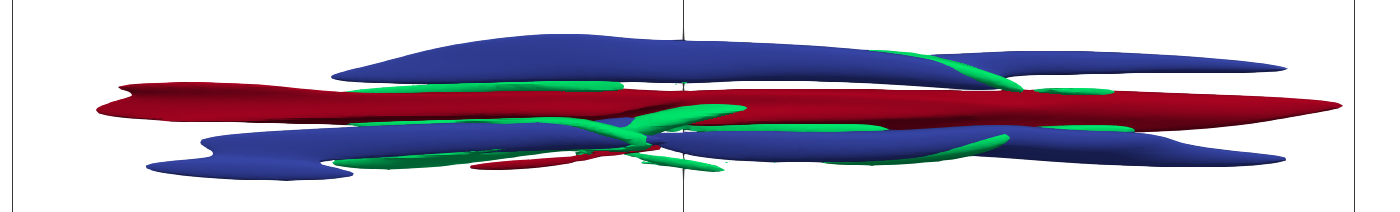}\\
    (c)
    \caption{Three-dimensional top view of the perturbation perturbation velocity field reference trajectory at (a) $t=100$, contours of $u=-8\times 10^{-3}$ (blue), $u=6\times 10^{-3}$ (red) and global $\lambda_2^*=-1\times 10^{-4}$, where $\lambda_2^*$ denotes the vortex identification criterion introduced by \cite{jeong_hussain_1995}. The distance between the vertical lines is $\Delta x =50$ (b) $t=280$, contours of $u=-5\times 10^{-2}$ (blue), $u=6\times 10^{-2}$ (red) and $\lambda_2^*=1\times 10^{-4}$. The distance between the vertical lines is $\Delta x =50$ and (c) $t=720$, contours of $u=-5\times 10^{-2}$ (blue), $u=6\times 10^{-2}$ (red) and global $\lambda_2^*=1\times 10^{-4}$. The distance between the vertical lines is $\Delta x =100$.}
    \label{fig:trajSnap}
\end{figure}
The minimal seed $M$ refers to the perturbation closest in kinetic energy to the laminar Blasius boundary layer flow, and able to trigger subcritical transition. This particular optimal condition was selected because it gives rise to a fully nonlinear trajectory relevant for the method tested. Moreover, it is uniquely defined by the parameters for the optimisation algorithm, namely here the Reynolds number value $Re_{\delta_0^{*}}=240.458$. That value of $Re_{\delta}(x=0)$ is chosen to match previous works \citep{Cherubini:2011aa,vavaliaris2020optimal}, in  particular the original work by \cite{Cherubini:2011aa} where the non-dimensionalisation differs from the present one. Note that in parallel flows the Reynolds number entirely defines the dynamical system, however in spatially developing flows the Reynolds number is intrinsically linked to the streamwise coordinate. Consequently, the minimal seed is conditioned by the range of Reynolds numbers (streamwise distances) allowed for in the time evolution of the perturbations. This results in the minimal seed being dependent on the inlet Reynolds number, on the length of the computational domain and on the optimization time \cite{vavaliaris2020optimal,beneitez2020modeling}. In \cite{vavaliaris2020optimal} the chosen optimization time is $T_{opt}=400$ and the computational domain length $L_x=500$. $M$ is computed iteratively using the nonlinear adjoint-based optimization framework  of \citet{rabin2012triggering}, \cite{Kerswell:2018aa}. The maximised objective function is the energy gain at a given time $T_{opt}$, $G(T_{opt})=E(T_{opt})/E(0)$ where $E(t)$ is the perturbation kinetic energy at time $t$. The optimization framework follows \cite{foures2013localization} and is based on the implementation into the open-source solver Nek5000 originally implemented by \cite{rinaldi2019vanishing}. The optimal state determined for a near-to-threshold initial energy $E_0$ was bisected using an edge tracking algorithm \citep{itano2001dynamics,Skufca2006}, so that the computed trajectory approximates well an edge trajectory for $t\le 800$, the bracketing trajectories differing by less than 2$\%$ in the observable used for edge tracking. This property is crucial for the stability study: initialising the base flow for the OTD analysis from outside the edge manifold would possibly result in a different transition scenario, as reported e.g. in \cite{Cherubini:2011aa}. Although the investigation in \cite{beneitez2019edge} warned against the possible interference between edge trajectories and unstable Tollmien-Schlichting waves over timescales $\mathcal{O}(10^{4})$, no such phenomenon will be encountered with the present set-up, since the considered observation time is $\mathcal{O}(10^{3})$.

A state portrait is shown in Fig.~\ref{fig:statePortrait_ref}, based on the three global quantities already used in previous studies \citep{Duguet2012,beneitez2019edge,beneitez2020modeling}.
\begin{eqnarray}
\Omega_x=(\delta_0^*/\delta)^{\frac{1}{2}}\left(\frac{1}{V}\int_V|\omega_x|^2\mathrm dv\right)^{\frac{1}{2}},
\label{eq:observables1}\\
\Omega_y=(\delta_0^*/\delta)^{\frac{1}{2}}\left(\frac{1}{V}\int_V|\omega_y|^2\mathrm  dv\right)^{\frac{1}{2}},
\label{eq:observables2}\\
W=(\delta_0^*/\delta)^{\frac{3}{2}}\left(\frac{1}{V}\int_V|w|^2\mathrm dv\right)^{\frac{1}{2}},
\label{eq:observables3}
\end{eqnarray} 
The quantities $\omega_x$ and $\omega_y$ are the streamwise and wall-normal perturbation vorticity components, respectively, and the integration is carried over the computation domain of volume $V$. The prefactors in powers of $(\delta_0^*/\delta)$ make use of the value of the boundary layer thickness evaluated at the center of mass, see \cite{Duguet2012}. In Fig.~\ref{fig:statePortrait_ref}, the edge trajectory is highlighted using a thicker (green) line, and the time interval $t\in[0,800]$ considered in this study is highlighted using a thicker green line (with equispaced dots every 50 time units). The thinner lines in red and blue correspond to trajectories closely  bracketing the edge trajectory.

It is useful to recall the main features of the unsteady base flow reported by \cite{vavaliaris2020optimal}. For early times $t \le 60$ the dynamics is dominated by a three-dimensional version of the Orr mechanism \citep{vavaliaris2020optimal}, where vortical disturbances initially tilted against the mean shear progressively untilt as time increases. For $60 \le t \le 200$ the lift-up mechanism takes over and a pair of streamwise streaks forms. Both mechanisms are known to be non-modal, the stronger energy amplification being associated with the lift-up \citep{schmid2001stability}. For $t \ge 200$ the energy growth slows down.  Snapshots of the velocity field along the optimal edge trajectory are shown in Figs. \ref{fig:trajSnap} at times $t=$100, 280 and 720. From $t \ge 100$ onwards, the edge trajectory consists of a localised pair of high- and low-speed streaks \citep{beneitez2019edge} with an undulation linked to oblique waves. It experiences a couple of streak-switching events around $t\approx 500$ and $t \approx 700$. The streaks elongate with time but remain always localised in the streamwise and spanwise direction. By construction, typical infinitesimal perturbations of this unstable flow field will make it evolve either towards an incipient turbulent spot or towards the laminar state. It is precisely their state space location on the verge of bypass transition that makes edge trajectories a relevant choice as a base flow \citep{khapko2016edge}. Imposing an optimality condition has the advantage of making the current trajectory well-defined.

\section{Results}

This section is devoted to the analysis of the stability properties of the optimal trajectory described in Section 3 using $r=8$ OTD modes. The choice of 8 modes aims at producing the largest possible subspace while keeping the simulations computationally feasible. The cost of each OTD mode is comparable to an additional DNS to be run in parallel to the original base flow. Moreover, the choice of number of modes is comparable with previous simulations of similar scale \citep{babaee2016minimization}. We restrain our study to the time interval $t\in [0,800]$. 
\subsection{Finite-time stability analysis}

\subsubsection{Instantaneous growth rates \label{sec:instantaneous}}

\begin{figure}
    \centering
    \includegraphics[width=0.8\textwidth]{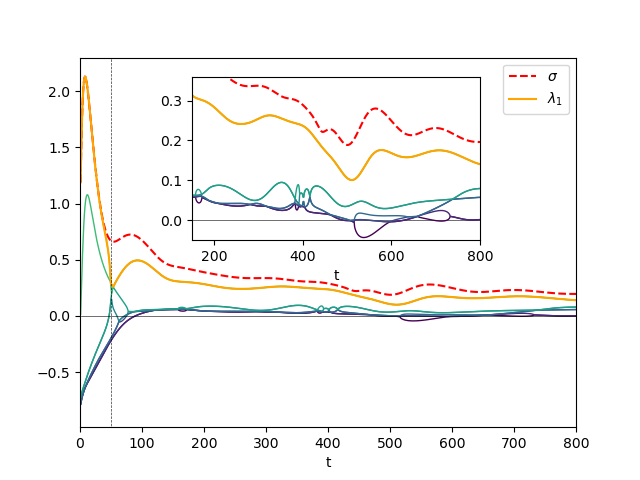}
    \caption{Real part of the instantaneous eigenvalues $\operatorname{Re}(\lambda_i),~i=1,..,8$ (solid lines) and instantaneous numerical abscissa $\sigma$ (dashed red) of the reduced operator ${\bm L}_r$, plotted versus time. $\lambda_1$ is highlighted (orange line). Vertical dashed line indicates $t=50$.}
    \label{fig:insteigs_short}
\end{figure}

\begin{figure}
    \centering
     \includegraphics[width=0.7\textwidth]{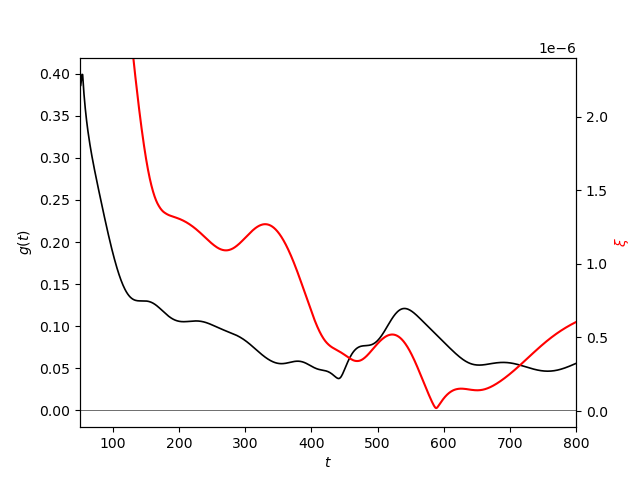}
    \caption{Quantifier $g(t)$ of the non-normality in the reduced system (black). Norm of the vector formed by the time derivatives of the observables used in the state portrait of Fig. \ref{fig:statePortrait_ref}: $d\Omega_x/dt$, $d\Omega_y/dt$ and $dW/dt$ as defined in Eq. \eqref{eq:roc} (red). The figure starts at $t=50$.}
    \label{fig:g}
\end{figure}

We begin by reporting the real part of the instantaneous eigenvalues $\lambda_i(t),~i=1,..,r$, computed over the whole trajectory. They are shown together with the instantaneous numerical abscissa $\sigma(t)$ versus time in Fig.~\ref{fig:insteigs_short}. The gap $g(t)=\sigma(t) - \operatorname{Re}(\lambda_1)$, which quantifies the instantaneous non-normality of the reduced operator, is displayed as a black line in Fig. \ref{fig:g}. These quantities have all been defined in Section \ref{sec:eigenmodes}.

The time series of these instantaneous growth rates can be grossly divided into two phases. In the initial phase for $t \lesssim$ 100, the two leading growth rates vary rapidly in time while the others are all negative. In a second phase starting at $t \approx$ 100, $\operatorname{Re}(\lambda_1)$ dominates in the range $0.2--0.3$, with a slight decaying trend as time increases. All other eigenvalues remain close to zero in real part, never exceeding 0.1. A quick glance at the state portrait in Fig. \ref{fig:statePortrait_ref} suggests that this second phase corresponds to a clear slowdown of the dynamics of the base flow itself.
If the dynamics is \emph{quasi-steady}, it is expected that the stability properties of the edge trajectory mimic qualitatively the
stability properties of steady/travelling edge states reported in other shear flow studies: one large dominating unstable eigenvalue, representing a strong instability in a direction transverse to the edge manifold, associated with many other eigenvalues of lesser magnitude responsible for the slow chaotic fluctuations \emph{within} the edge manifold \citep{duguet2008transition}. This expectation is largely confirmed by  Fig.~\ref{fig:insteigs_short} for $t \ge$ 100.

A finer analysis of the fluctuations of the growth rates is possible both in the initial and the quasi-steady phases. This is achieved by focusing on the gap $g(t)$, interpreted as a measure of instantaneous non-normality within the OTD subspace. For the initial times $t \lesssim$ 50, $\sigma=\operatorname{Re}(\lambda_1)=\operatorname{Re}(\lambda_2)>0$. After $t=50$, the gap $g$ rises from zero to a maximum of about 0.4. It later decreases to smaller values of $\approx$ 0.1. As for the other $\lambda_i$'s, they are all negative at $t=0$ but grow at the same pace and cross zero at $t \approx$ 100. At later times, all instantaneous growth rates stabilise, while $\operatorname{Re}(\lambda_1)$ decreases gently in a non-monotonic manner, and $g(t)$ oscillates around low values $\approx$ 0.05--0.1. The fact that the peak of $g(t)$ occurs before $t=$100 is consistent with the reported occurrence of purely non-normal Orr and lift-up mechanisms along the edge trajectory for these times \citep{vavaliaris2020optimal}. The sensitivity of the edge trajectory appears high where the edge trajectory also experiences strong non-normal amplification. However, the fact that $g \approx 0$, i.e. $\sigma=\lambda_1$ at the earliest times $t \le$50 may be wrongly attributed to a lack of non-normal potential of $\bm L(t)$. To start with, this is a property of the instantaneous reduced operator $\bm L_r(t)$ computed for a given value of $r$, not necessarily of the full operator $\bm L(t)$. The reverse is yet true: non-normal features of the reduced-order operator $\bm L_r(t)$ carry over to $\bm L(t)$. Moreover, after trying several different initialisations this result was found to depend crucially on the choice of the OTD basis for $t=0$, at least over early times $t\le 50$. This makes it difficult to draw general conclusions for short enough times, consistently with the study of \cite{babaee2017}. This is possibly confirmed by the very transient behaviour of the eigenvalues $\lambda_2$ to $\lambda_8$. From $t\approx 50$ on, the non-normal potential within the OTD subspace is high again as expected, judging from the large values of $g(t)$, and transient effects due to the initialisation of the OTD modes can be neglected.

A peak at $t \approx 60$, and a smaller one at $t\approx 100$, are evident in the data for $\sigma(t)$ in figure \ref{fig:insteigs_short}(a). These times are perfectly consistent with the occurrence of both the Orr and the lift-up mechanisms described in \cite{vavaliaris2020optimal}. Two additional bumps for both $g(t)$ and $\sigma(t)$ can also be seen at $t\approx 550$ and $t \approx 720$. According to \cite{vavaliaris2020optimal}, these two times correspond to \emph{streak switching events}. This suggests that streak-switching events, themselves an inherent part of the self-sustained mechanism \citep{khapko2013localized,beneitez2019edge}, are linked to stronger non-normality than the rest of the edge trajectory.

Eventually, Fig. \ref{fig:g} also shows the norm of the time-derivatives of the three observables $\Omega_x$, $\Omega_y$ and $W$ used in Fig. \ref{fig:statePortrait_ref}. This quantity is defined as $\xi$ via
\begin{equation}
    \xi(t) = \sqrt{\left(\frac{d\Omega_x}{dt}\right)^2+\left(\frac{d\Omega_y}{dt}\right)^2+C\left(\frac{dW}{dt}\right)^2}, \label{eq:roc}
\end{equation}
where $C$ is a unity-valued constant ensuring the correct dimensionality.
It is plotted in fig. \ref{fig:g} in connection with the time evolution of $g(t)$. We analyse now these quantities by considering  consecutive sub-intervals of the edge trajectory starting from the minimal seed:  (i) $t\in[0,60]$ (Orr mechanism in the base flow) corresponds to a very rapid evolution of the observables $\xi(t)$. The OTD modes, however, take time to catch up with non-normality until $t\approx 80$, as shown by $g(t)$. (ii) $t\in[60,200]$ corresponds to the lift-up in the base flow effect associated with non-normal growth. Here, $g(t)$ appears largest for $t\approx 80$ and decreases rapidly until $t\approx 130$, where a change in the slope of $g(t)$ can be noticed. $\xi (t)$ mirrors this behaviour, suggesting that non-normality is decreasing as the lift-up of the base flow ends. (iii) The trajectory has reached the relative attractor past $t\geq 200$. In this stage we observe that the slow-down of the dynamics indicated by $\xi(t)$ corresponds to higher values of $g(t)$, and \emph{vice versa}. This can be seen in the intervals $t\in [300,400]$, where the dip in $g(t)$ corresponds to a peak in $\xi(t)$, and in $t\in [500,600]$, where an increase in $g(t)$ corresponds to a dip in $\xi(t)$.

\subsubsection{Characterisation as an outer mode instability}

\begin{figure}
    \centering
    \includegraphics[width=0.49\textwidth]{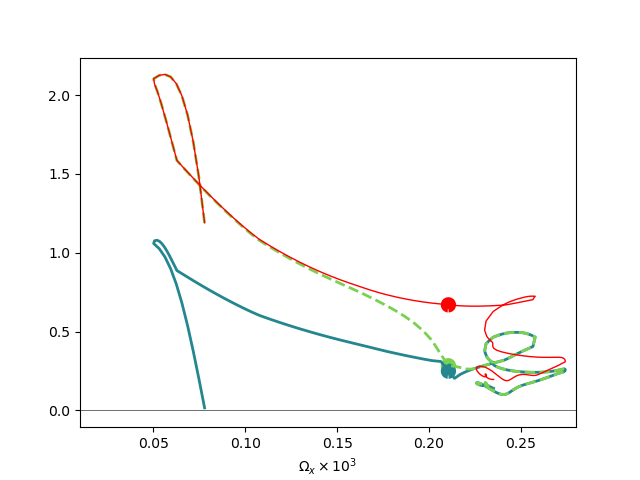}
    \includegraphics[width=0.49\textwidth]{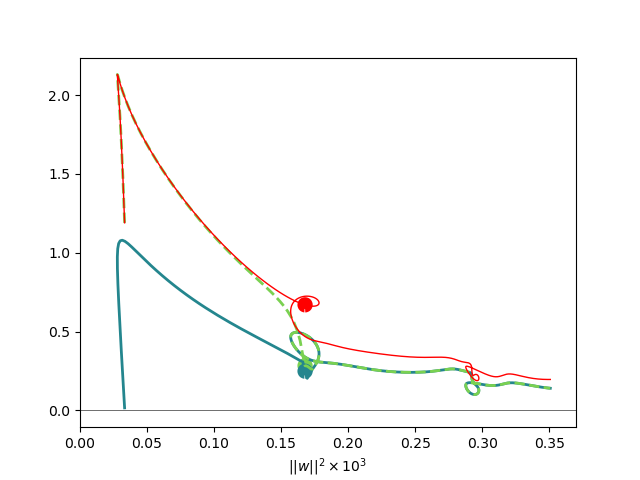}
     \caption{(a) Green lines: $\operatorname{Re}(\lambda_{1,2})$ vs $\Omega_x$, red line $\sigma$. The dots denote $t=50$, the trajectory starts from the values on the left of the figure. (b) Right: \textit{idem} for $\operatorname{Re}(\lambda_{1,2})$ vs $||w||^2$, the volume average spanwise velocity.} 
    \label{fig:insteigs_shortbis}
\end{figure}

In the original study on streak breakdown by \citet{vaughan2011stability}, where the base flow consists of a quasi-steady localised streak rather than a time-dependent one, a distinction was made between two types of modes. The main criterion is the wall-normal position of the energy of each mode with respect to the location of the critical layer, the latter being known from inviscid analysis. The modes with a critical layer close to the wall (such as Orr-Sommerfeld modes) are denoted as inner modes, while those with a critical layer in the free-stream are denoted as outer modes. Another characterisation of the inner \emph{vs.} outer mode distinction, also suggested by by \citet{vaughan2011stability}, relies on the relation between the growth rate of the mode and the streak amplitude. Although the present context differs, notably because of the unsteady aspect of the streaks, such a characterisation can also be applied to the modes determined by our method.
Fig. \ref{fig:insteigs_shortbis} shows the values of the two largest instantaneous eigenvalues $\operatorname{Re}(\lambda_{1,2})$ plotted \emph{vs.} $\Omega_x$, in Fig. \ref{fig:insteigs_shortbis} (a), and \emph{vs.} the volume average energy in the spanwise direction $||w||^{2}$. These quantities are used as a proxy for the instantaneous amplitude of the streaky edge state.  It can be directly compared to Fig. 7 from \citet{vaughan2011stability}, where the growth rate is plotted versus the streak amplitude (called $A_u$). The corresponding figure was used to define a classification of the instability mechanisms: \emph{inner mode} refers to an instability mode present for arbitrary small values of $A_u$, in contrast with \emph{outer modes} which are not found for vanishing streak amplitude. In the present case, positive growth rates are only found for non-vanishing values of the observable $\Omega_x \ge 0.05$. Interpreting $\Omega_x$ as an alternative definition of streak amplitude unambiguously indicates that the dominant instability of the edge state should be classified as an \emph{outer mode} instability.

\subsubsection{Finite-time Lyapunov exponents}

\begin{figure}
    \centering
    \begin{tabular}{cc}
       \includegraphics[width=0.495\textwidth]{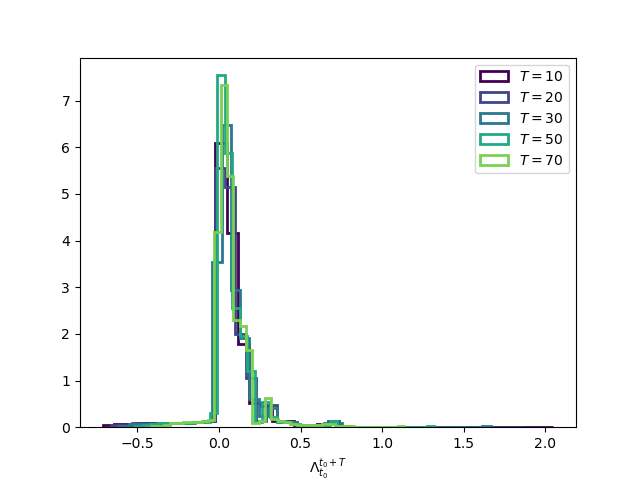}  & \includegraphics[width=0.495\textwidth]{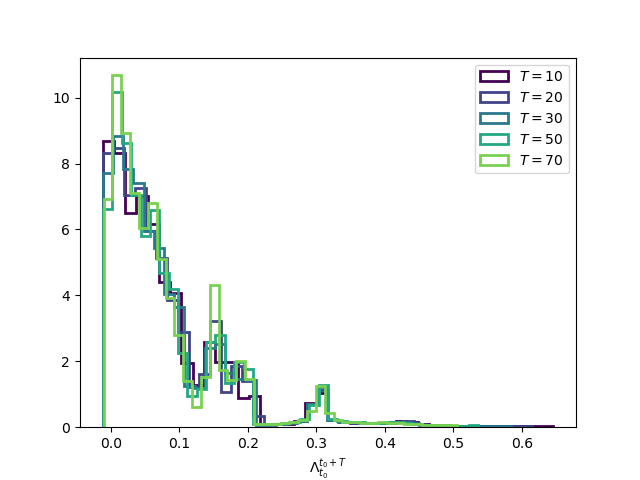}  \\
        (a) & (b)
    \end{tabular}
    \caption{Histograms of all $r=8$ reduced-order FTLEs $(\Lambda_{t_{0}}^{t_{0}+T})_i$ ($i=1,..,8$) sampled over various time window values of $t_0 \in [0,720]$ along the unsteady base flow trajectory. Horizon times $T=$10, 20, 30, 50 and 70. The histograms are normalised so that their integral is equal to 1. (a) $t_0 \in [0,720]$.  (b) $t_0 \in [100,720]$.}
    \label{fig:insteigs_short2}
\end{figure}

Fig.~\ref{fig:insteigs_short2}(a) shows distributions of FTLEs $(\Lambda_{t_{0}}^{t_{0}+T})_i$ ($i=1,..,8$) computed within the interval $t_0 \in [0,800]$. Fig.~\ref{fig:insteigs_short2}(b) is similar except that the values of $t_0$ are restrained to the sub-interval $t_0 \in [100,800]$. In both plots the time horizon $T$ takes increasing values from 10 to 70. Comparing the different values of $T$ essentially confirms the robustness of the FTLE distributions with respect to the time horizon. The reason why all FTLES from $i=$1 to 8 are reported together is the frequent change in the ordering of the growth rates, occurring every time an eigenvalue crossing takes place \citep{babaee2017}. The many negative occurrences in Fig.~\ref{fig:insteigs_short2}(a), as well as the largest occurences ($\ge$0.3) can be attributed to the choice of initial conditions for the OTD modes, including accidentally co-aligned disturbances. A potential improvement of the initial conditions could be the computation of the eigenvectors associated with the minimal seed, by assuming no time dependency. This would result in eigendirections already within the initial tangent space. Even though there is no guarantee that these directions will remain in the tangent space at later times, they can be expected to be physically relevant at least for the initial times. These occurrences indeed disappear entirely in Fig.~\ref{fig:insteigs_short2}(a) after the initial first 100 time units have been discarded, consistently with the results of Section \ref{sec:instantaneous}. Since the original bisection algorithm is essentially a shooting method \citep{itano2001dynamics}, we expect one of the FTLEs to be the signature of the instability of the edge manifold. In other words this FTLE is associated with an unstable direction pointing transversally to it. The other additional positive FTLEs have no choice but to be associated with the weak apparent unsteady dynamics taking place \emph{within} the relative attractor, rather than transversally to it. This conclusion is consistent with the results of Section \ref{sec:instantaneous}. The two peaks in Fig.~\ref{fig:insteigs_short2}(a), close to $0.15$ and $0.3$, correspond to a higher number of occurrences. They can be related respectively to the slow and fast separation of vortical disturbances, later to be shed from the main edge structure, see \cite{Duguet2012}.

\subsubsection{Local expansion rates}

\begin{figure}
    \centering
    \begin{tabular}{cc}
      \includegraphics[width=0.495\textwidth]{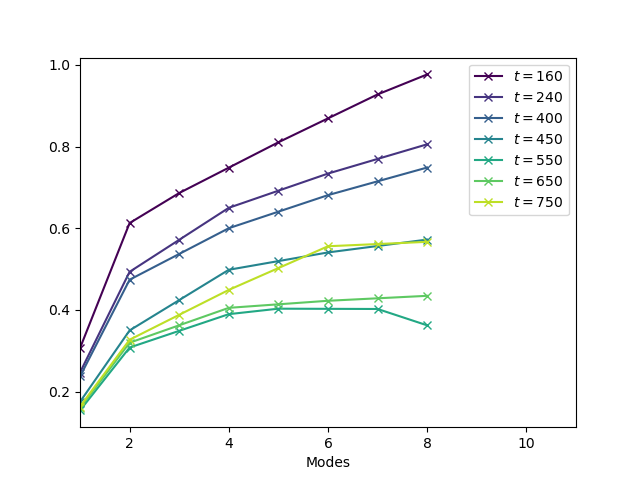}  & \includegraphics[width=0.495\textwidth]{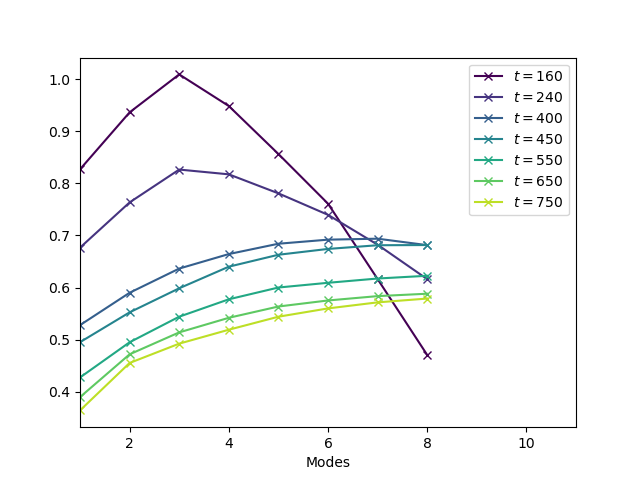} \\
      (a) & (b) 
    \end{tabular}
    \caption{(a) Cumulative sum $S_j(t)$ of $\operatorname{Re}(\lambda_i),~i=1,..,j$ at different times and for $j=1,..,8$ (b) Cumulative sum of $\Lambda_{t_{0}}^{t_{0}+T}$  at different times.}
    \label{fig:insteigs_short2bis}
\end{figure}

When dealing with proper attractors defined over unbounded times, it is common to estimate numerically its dimension. Among the different possible definitions, the Kaplan-Yorke dimension $D_{KY}$ is of interest, because it only requires the values of the leading Lyapunov exponents $\lambda_i$, once ranked in descending order $\lambda_1>\lambda_2>...>\lambda_r$. It is defined as
\begin{eqnarray} 
D_{KY}=j+\frac{S_j}{|\lambda_{j+1}|},
\label{eq:D_KY}
\end{eqnarray}
where $S_j$ the cumulative sum 
\begin{eqnarray} 
S_j=\sum_{i=1}^{j}\lambda_{i},
\label{eq:cumulative}
\end{eqnarray}
and $j$ is the only integer such $S_j>0$, but $S_{j+1}<0$. 
In the present case the long-time Lyapunov exponents $\lambda_1,...$. cannot be computed since the dynamics takes place over finite times. The above definition can however be generalised to finite-time problems by considering either the instantaneous or the finite-time exponents \citep{kuptsov2018lyapunov}. The current analysis is based on the sum $S_j$, rather than on the effective dimension $D_{KY}$ which can be constructed from $S_j$ in eq. \ref{eq:cumulative} only if $j$ is large enough. Indeed with the present value of $r=8$, there are not enough negative exponents to define $D_{KY}$ according to eq. \eqref{eq:D_KY}. Geometrically, $S_j(t)$ is understood as the instantaneous rate-of-change of the volume of an infinitesimal state space element defined in the corresponding $j$-dimensional subspace \citep{kuptsov2018lyapunov}. In Fig.~\ref{fig:insteigs_short2bis} we show the cumulative sum $S_j$(t) as a function of time, computed in two different ways. Fig.~\ref{fig:insteigs_short2bis}(a) has $S_j(t)$ based on the instantaneous growth rates $\operatorname{Re}(\lambda_i),~i=1,..,j$. Fig.~\ref{fig:insteigs_short2bis}(b) has $S_{t_0}^{t_0+T}$ based on the FTLEs $\Lambda_{t_{0}}^{t_{0}+T}$, which are computed over an entire time interval.

It is observed that, for $j\le 8$ and $t\le800$, both cumulative sums never become negative. This confirms that the instantaneous and the finite-time Kaplan-Yorke dimension of the underlying relative attractor are both strictly larger than $8$. Interestingly, $S_j$ decreases with $t_0$, up to $t=550$, for all $j$'s. For the last values of $t_0$ plotted, $S_j$ even eventually decreases with j, which suggests that instantaneous eigenvalues with negative growth rate start to contribute to the instantaneous/FTLE spectrum at later times. From a geometric point of view, the fact that $S_j$ stays always positive suggests that the volume of infinitesimal state space elements of the reduced $r$-dimensional space grows with time. This is in contrast with the full $n$-dimensional space where such a volume has to decrease, since the original dynamical system \eqref{eq:main} is dissipative. In other words, the present reduction, with the choice of $r=8$, does not incorporate enough dissipative modes, only active modes. Conducting a similar numerical experiment with much larger $r$ is as of today too demanding in terms of memory requirements, at least for the Blasius flow.

\subsubsection{Summary}

The main learnings from the OTD stability analysis restrained to $r$=8 modes are the following: the dominant edge instability qualifies an outer mode mechanism linked with the wall-normal vorticity of the localised streak. Past the initial 50 time units where the analysis depends on the initialisation of the modes, several mechanisms can be identified from the three peaks in the FTLE spectrum. The dominant instability corresponds to an instability transverse to the edge manifold, while the others correspond to the slow variability of the edge trajectory itself: the dynamics of the perturbations mimic the dynamics inherent to the base flow itself, including the streak phenomenon. The local dimension of the tangent space exceeds the value of $r$=8. Finally, we observe that the non-normal amplification of disturbances increases when the change of the base flow in time becomes slower and \emph{vice versa}.

\subsection{Modal structures}

Beyond global indicators characterising the tangent dynamics, a description of the modal structures in physical space is required. We recall (see Subsection \ref{sec:eigenmodes}) that the flow fields visualised correspond to the real part of the vectors ${\bm u}_i^{\lambda}$ defined in eq. \ref{eq:Ulambda}. Two-dimensional visualisations are shown for two different times, namely $t=280$ and 720. The modes come in complex conjugate pairs for the considered times, therefore we only display here every other mode among the computed ones. The velocity field of the base flow at these two times, selected along the edge trajectory after the initial transient, is shown in Fig. \ref{fig:trajSnap}. It consists of a wiggly finite-length streak flanked with shorter streamwise vortices. At these two times, both snapshots are comparable, the main differences being the longer streamwise extent together (of about 500$\delta_0^*$) with a spanwise narrower structure of extent 40$\delta_0^*$ at the later time. Taking into account the dynamics of the base flow near these two times enriches the description. Near $t=280$ the formation of streaks by the lift-up mechanism is almost mature \citep{vavaliaris2020optimal} and the dynamics relaxes towards quasi-steady motion. By contrast, in the time units following $t=720$, low and high-speed streak are on the verge of exchanging their spanwise position.

\begin{figure}
    \centering
    \includegraphics[width=0.8\textwidth]{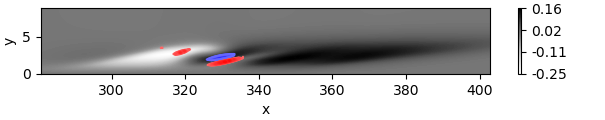}
    \includegraphics[width=0.8\textwidth]{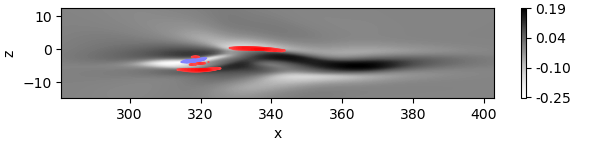}
    \includegraphics[width=0.8\textwidth]{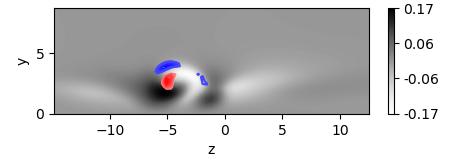}
    \caption{$t=280$, $u^{\lambda}_1$. Two-dimensional greyscale planes of the perturbation velocity of the edge trajectory with arbitrary amplitude, contour lines of $\omega_k\ k=\{x,y,z\}$, normal to the corresponding plane for each $\left\{u_i^{\lambda}\right\}$, corresponding to 40\% to 100\% of its maximum value. For each mode, from top to bottom the planes displayed are $z=-4$, $y=2.5$ and $x=325$.}
    \label{fig:OTD1_t280}
\end{figure}

\begin{figure}
    \centering
    \includegraphics[width=0.8\textwidth]{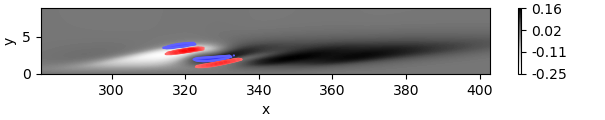}
    \includegraphics[width=0.8\textwidth]{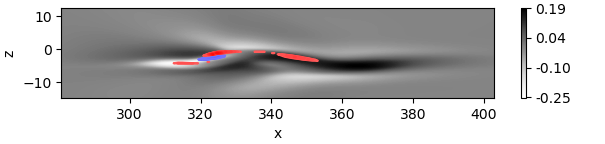}
    \includegraphics[width=0.8\textwidth]{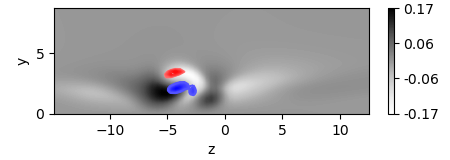}
    \caption{$t=280$. Mode $u^{\lambda}_3$. Same as Fig.\ref{fig:OTD1_t280}.}
    \label{fig:OTD3_t280}
\end{figure}

\begin{figure}
\centering
    \includegraphics[width=0.8\textwidth]{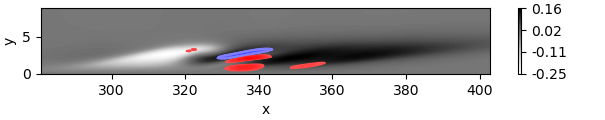}
    \includegraphics[width=0.8\textwidth]{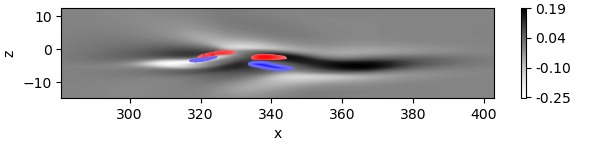}
    \includegraphics[width=0.8\textwidth]{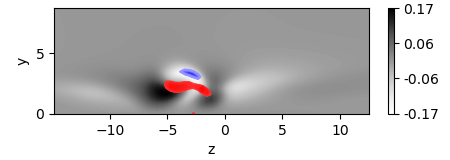}
    \caption{$t=280$. Mode $u^{\lambda}_5$. Same as Fig.\ref{fig:OTD1_t280}.}
    \label{fig:OTD5_t280}
\end{figure}

\begin{figure}
    \centering
    \includegraphics[width=0.8\textwidth]{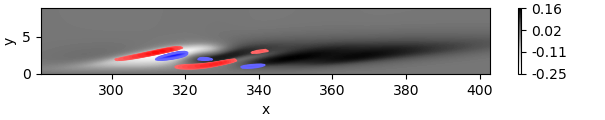}
    \includegraphics[width=0.8\textwidth]{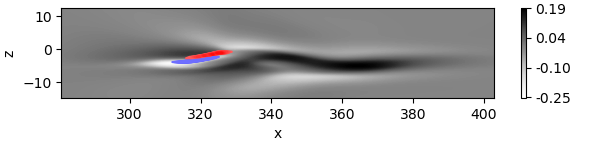}
    \includegraphics[width=0.8\textwidth]{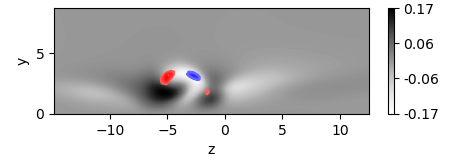}
    \caption{$t=280$. Mode $u^{\lambda}_8$. Same as Fig.\ref{fig:OTD1_t280}.}
    \label{fig:OTD8_t280}
\end{figure}

\begin{figure}
    \centering
    \begin{tabular}{cc}
    \includegraphics[width=0.495\textwidth]{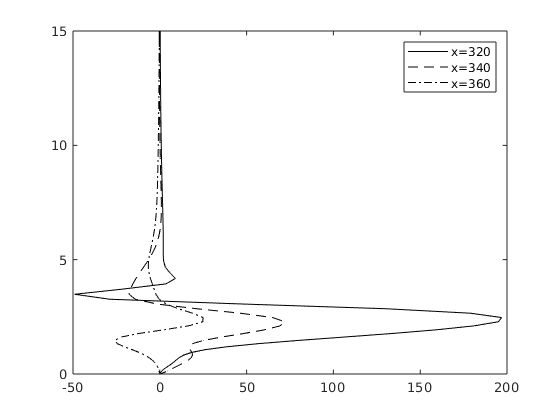} & \includegraphics[width=0.495\textwidth]{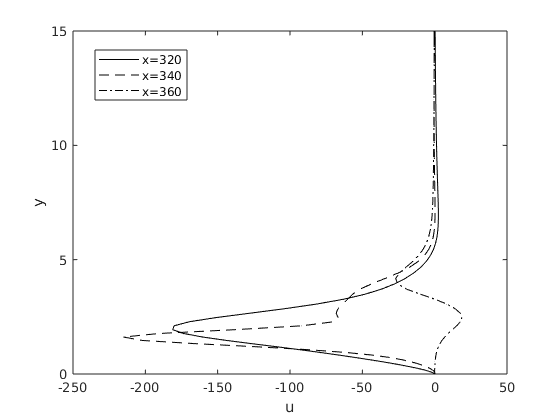} \\  
       (a) & (b) \\
    \includegraphics[width=0.495\textwidth]{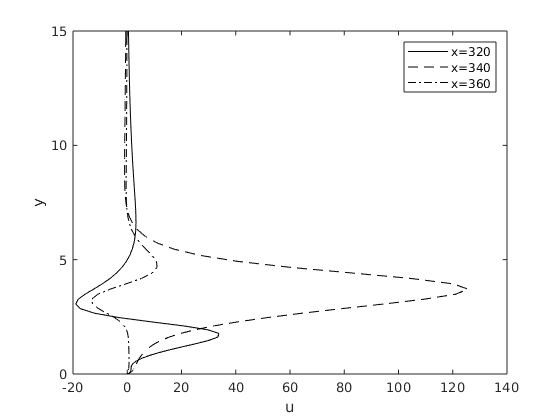} &
    \includegraphics[width=0.495\textwidth]{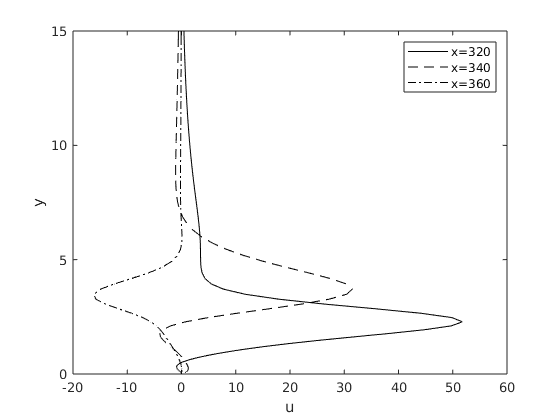} \\
    (c) & (d)
    \end{tabular}
    \caption{Streamwise velocity profiles of $u_1^{\lambda}$ at $t=280$ corresponding to positions $x=320$ (solid), $x=340$ (dashed), $x=360$ (dot-dashed) for (a) $z=-4$ (b) $z=-2$ (c) $z=2$ (d) $z=4$.}

    \label{fig:uprofs_t280}
\end{figure}
\begin{figure}
    \centering
    \begin{tabular}{cc}
       \includegraphics[width=0.495\textwidth]{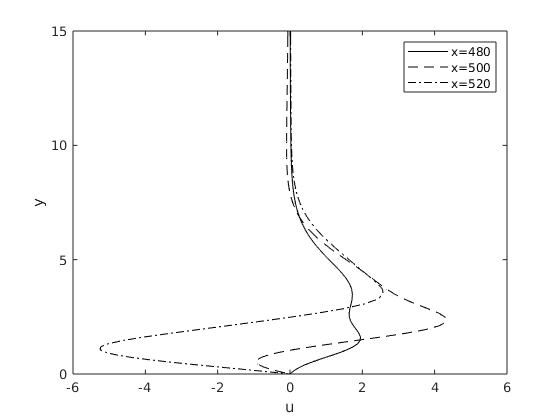} & \includegraphics[width=0.495\textwidth]{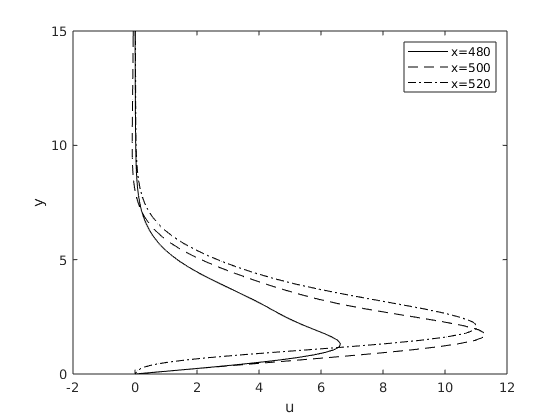}   \\
       (a)  & (b) \\
       \includegraphics[width=0.495\textwidth]{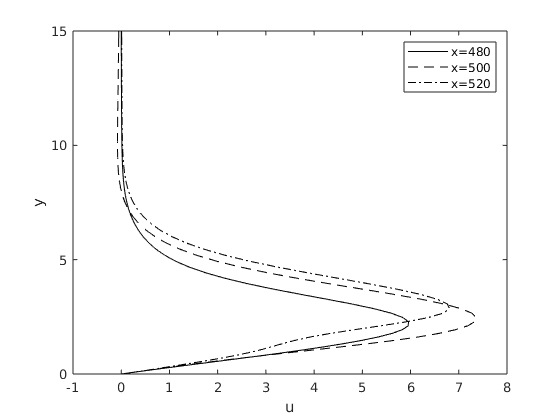} &
       \includegraphics[width=0.495\textwidth]{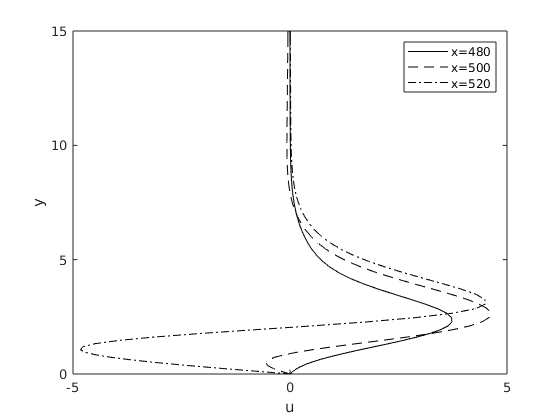}\\
       (c) & (d)
    \end{tabular}
    \caption{Streamwise velocity profiles of $u_1^{\lambda}$ at $t=720$ corresponding to positions $x=480$ (solid), $x=500$ (dashed), $x=520$ (dot-dashed) for (a) $z=-4$ (b) $z=-2$ (c) $z=2$ (d) $z=4$.}
    \label{fig:uprofs_t720}
\end{figure}

The instantaneous eigenmodes for $t=280$ are first shown in Figs.~ \ref{fig:OTD1_t280}--\ref{fig:OTD8_t280}. The representation, inspired by the experimental figures of \cite{balamurugan2017experiments}, is based on a pseudocolor plot of the streamwise velocity perturbation for the reference trajectory, overlapped with lines indicating 40\%-100\% of the maximum range of the vorticity normal to the planes at $z=-4$, $y=2.5$ and $x=325$. The planes are selected to intersect relevant regions of the main structure. We describe now the observed flow structures. The present method as well as the underlying modal decomposition are new in fluid mechanics
apart from \cite{babaee2016minimization}. Therefore for pedagogic reasons we chose to display the flow fields of every computed instantaneous eigenmode, omitting the redundant conjugate modes.

For $t=280$, the spatial structure of each of the 8 leading OTD modes superimposes well with the active part of the main structure, which consists of a \emph{sinuous} streak of finite length. As a consequence the OTD modes inherit this sinuous structure.
Importantly, no spatial symmetry has been imposed neither on the base flow nor on the disturbances modes. This differs from the classical study of \cite{andersson2001breakdown} where the base flow has no streamwise dependence. The long-standing question about the symmetries of the leading eigenmodes, namely whether they are symmetric with respect to the plane $z=0$ (sinuous) or antisymmetric (varicose), becomes irrelevant here. In particular the varicose symmetry, which is consistent with the formation of hairpin vortices, is not characteristic of any of the modes investigated. The classical conclusion of \cite{andersson2001breakdown}, namely that the sinuous instability of streaks is the most unstable mechanism of paramount importance for streak breakdown, remains valid. Further visualisation of the modes at $t=280$ highlights the shear layers in the flow, visible in the $xy$ plane. The $xz$-plane shows that most of the activity of the mode is located within the active core of the streak and its upstream tail. The $yz$-plane confirms the localisation of the mode on the top shear layer. For all modes, energy is located mostly within the active core or upstream of it. This is in line with the former observation that secondary structures shed downstream of the edge state are \emph{not} key ingredients of the self-sustained cycle \citep{Duguet2012}. Streamwise velocity profiles for the instantaneous eigenmodes are shown in Fig.~\ref{fig:uprofs_t280}. They suggest robust localisation close to the edge of the boundary layer. In all subfigures in Fig.~\ref{fig:uprofs_t280}, the $y$-location for the largest amplitude of the streaks is displaced towards larger values with increasing $x$: the head of the streaks characterising the edge trajectory appears tilted upwards. This is again consistent with the description of outer mode instability in \cite{vaughan2011stability}. The relevance of this region is furthermore consistent with the interpretation in \cite{hack2014streak}, where streak instability proceeds via outer modes localised near the edge of the boundary layer. As for the differences between the different modes $u_1^{\lambda},...,u_8^{\lambda}$, at $t=280$ they are not very pronounced yet. Only $u_1^{\lambda}$ stands out through a less pronounced tail of streamwise vorticity at the upstream edge. It was checked that perturbing the edge trajectory at $t=280$ by $u_1^{\lambda}$, with amplitude $\pm 10^{-4}$, leads either to a turbulent flow or to relaminarisation. This confirms that this eigendirection is transverse to the edge manifold at the considered time.

\begin{figure}
    \centering
    \includegraphics[width=0.8\textwidth]{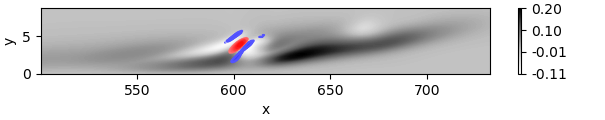}
    \includegraphics[width=0.8\textwidth]{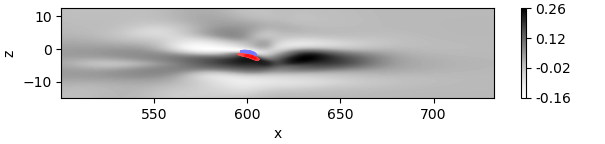}
    \includegraphics[width=0.8\textwidth]{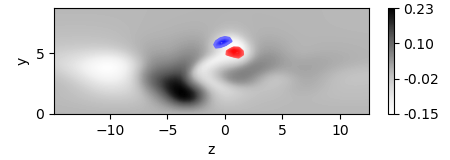}
    \caption{$t=720$, $u^{\lambda}_1$. Two-dimensional greyscale planes of the perturbation velocity of the edge trajectory with arbitrary amplitude, contour lines of $\omega_k\ k=\{x,y,z\}$, normal to the corresponding plane for each $\left\{u_i^{\lambda}\right\}$, corresponding to 40\% to 100\% of its maximum value. For each mode, from top to bottom the planes displayed are $z=-1.6$, $y=2.5$ and $x=610$.}
    \label{fig:OTD1_t720}
\end{figure}
\begin{figure}
    \centering
    \includegraphics[width=0.8\textwidth]{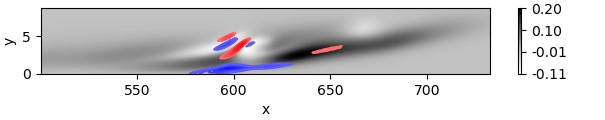}
    \includegraphics[width=0.8\textwidth]{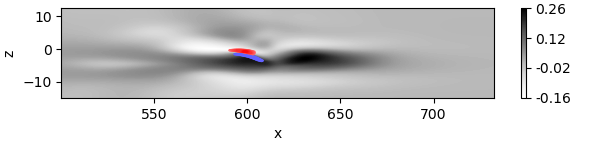}
    \includegraphics[width=0.8\textwidth]{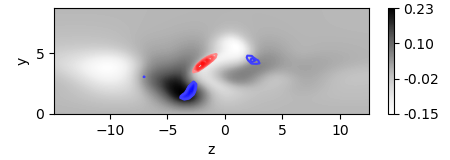}
    \caption{$t=720$. Mode $u^{\lambda}_3$. Same as Fig.\ref{fig:OTD1_t720}.}
    \label{fig:OTD3_t720}
\end{figure}
\begin{figure}
    \centering
    \includegraphics[width=0.8\textwidth]{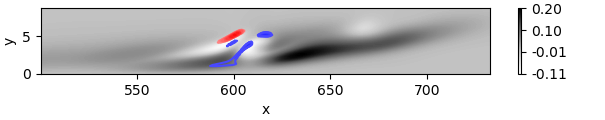}
    \includegraphics[width=0.8\textwidth]{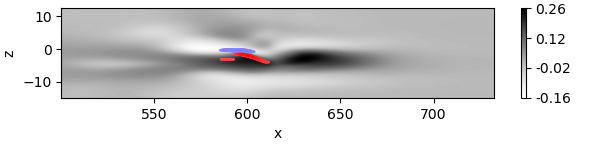}
    \includegraphics[width=0.8\textwidth]{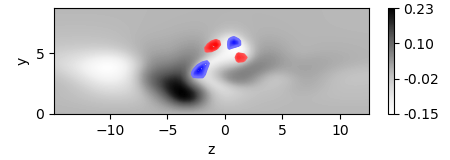}
    \caption{$t=720$. Mode $u^{\lambda}_5$. Same as Fig.\ref{fig:OTD1_t720}.}
    \label{fig:OTD5_t720}
\end{figure}
\begin{figure}
    \centering
    \includegraphics[width=0.8\textwidth]{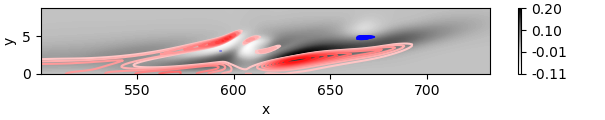}
    \includegraphics[width=0.8\textwidth]{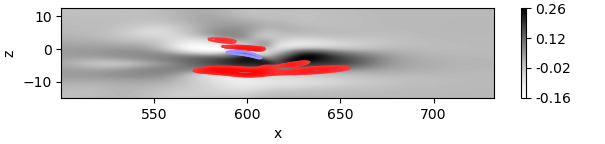}
    \includegraphics[width=0.8\textwidth]{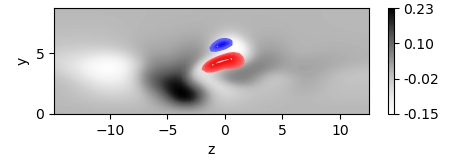}
    \caption{$t=720$. Mode $u^{\lambda}_8$. Same as Fig.\ref{fig:OTD1_t720}. Note that the large red structure in the $xy$-plane is located in the region where a new streak will be spanned \citep{vavaliaris2020optimal} and the supplementary material therein.}
    \label{fig:OTD8_t720}
\end{figure}

Most features discussed above are also attested at a later time $t=720$, just before streak switching takes place. There are however noticeable differences. At $t=720$, all the instantaneous eigenvalues are still positive, with $\lambda_1$ strictly larger than the other eigenvalues and $\lambda_8$ closer to zero. The leading OTD mode $u_1^{\lambda}$ is still similar in shape to $u_3^{\lambda}$ and $u_5^{\lambda}$, while $u_8^{\lambda}$ clearly displays a different structure. $u^{\lambda}_1$ displays strong activity at the edge of the boundary layer, upstream of the active core, strictly above the corresponding shear layer of the base flow (it is most visible on the streamwise velocity component). More noticeable is the fact that the modal structures are lifted towards the edge of the boundary layer, see e.g. the $xy$ plane of Fig.~\ref{fig:OTD1_t720}(a) for $u^{\lambda}_1$. The vortical structures associated with this mode form a larger angle with the wall than the base flow itself. It was again checked that the eigendirection $u^{\lambda}_1$ is transverse to the edge manifold at the considered time.
The structures highlighted in $u^{\lambda}_8$ are of particular interest. They correspond to the region where a new high-speed streak is in the process of being spanned (see the supplementary material in \cite{vavaliaris2020optimal} for further evidence). The corresponding OTD mode(s) should hence not only be understood as the manifestation of an instability of a simple instability-free base flow, instead it can be interpreted as precursor(s) of events that will anyway occur along the edge trajectory. The positive FTLEs associated with the corresponding instantaneous eigenmode are a signature of short-term unpredictability, they quantify the temporal volatility of the streak switching phenomenon.

\begin{figure}
    \centering
    \vspace{5mm}
    \includegraphics[width=0.9\textwidth]{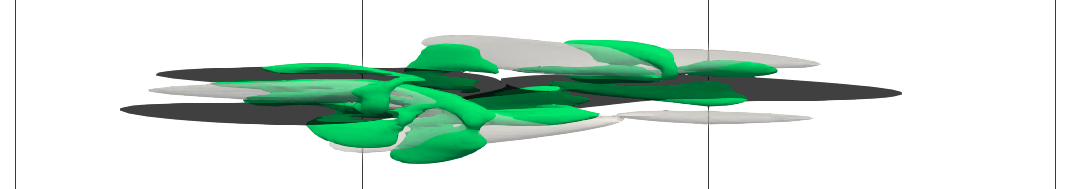}\\
    (a)\vspace{5mm}\\
    \includegraphics[width=0.9\textwidth]{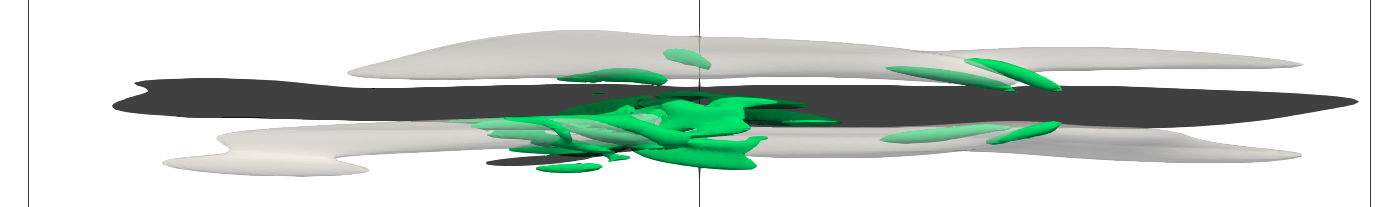}\\
    (b)
    \caption{Three-dimensional top view of the leading instantaneous eigenmode ${\bm u}_1^{\lambda}$ with arbitrary amplitude, superimposed on the edge trajectory from Fig.~\ref{fig:trajSnap} (a) $t=280$, contours of $u=-5\times 10^{-2}$ (white), $u=6\times 10^{-2}$ (black) for the same snapshot as in Fig. \ref{fig:trajSnap} and $\lambda_2=-0.29\%$ of its maximum value (green) for the leading OTD mode. (b) $t=720$, contours of $u=-5\times 10^{-2}$ (white), $u=6\times 10^{-2}$ (black) for the same snapshot as in Fig. \ref{fig:trajSnap} and $\lambda_2=-0.47\%$ (green) of its maximum value for the leading OTD mode.} 
    \label{fig:3DOTD}
\end{figure}

\begin{figure}
    \centering
    \includegraphics[width=0.9\textwidth]{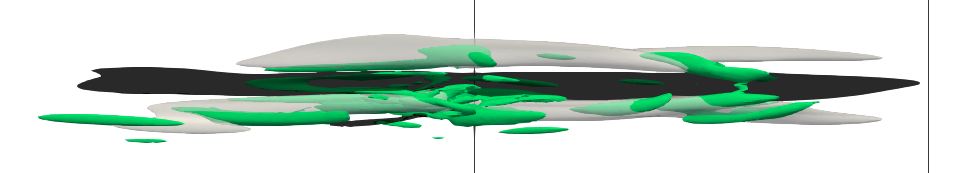}
    \caption{Three-dimensional top view of the eighth instantaneous eigenmode ${\bm u}_8^{\lambda}$, with arbitrary amplitude, superimposed on the edge trajectory from Fig.~\ref{fig:trajSnap} $t=720$, contours of $u=-5\times 10^{-2}$ (white), $u=6\times 10^{-2}$ (black) for the same snapshot as in Fig. \ref{fig:trajSnap} and $\lambda_2=-2.3\%$ (green) of its maximum value.}
    \label{fig:3DOTD8}
\end{figure}

Further strengthening the discussion above, Fig. \ref{fig:3DOTD} shows the same snapshots as in Fig. \ref{fig:trajSnap} superimposed now with contours of $\lambda_2$ for the leading OTD mode. It can be seen in both Fig. \ref{fig:3DOTD}(a) and (b) that the instability mode is mostly localised within the edge  structure. The localisation within the active core is even clearer in Fig. \ref{fig:3DOTD}(b). Furthermore, Fig. \ref{fig:3DOTD8} shows greater localisation in the side where a new high-speed streak is to be generated. 

Some elements of this analysis could have been anticipated. The OTD framework, in line with the whole concept of Lyapunov analysis, is a generalisation of modal stability analysis to arbitrarily unsteady base flows. Non-normal features can be captured provided an insightful initialisation of the OTD modes, yet these features are not expected to persist over longer time horizons, e.g. those involved in the evaluation of FTLEs. However, the Orr as well as the lift-up mechanism, which dominate the dynamics at early times, are intrinsically non-normal mechanisms of finite duration. In principle, a large number of eigenvectors is needed to capture transient growth accurately. This explains why so many modes possess a similar structure.  This trend is aggravated by the fact that for small $r$, the captured non-normality is an estimate of the non-normality of the whole system.

If the description in terms of few OTD modes can seem irrelevant at the earliest times when non-normality dominates, the situation becomes tractable again with small $r$ as soon as the growth of the streaks slows down. The corresponding visualisations for $t=280$ and $t=720$ are displayed in Fig.~\ref{fig:OTD1_t280}--\ref{fig:OTD8_t280} and Fig.~\ref{fig:OTD1_t720}--\ref{fig:OTD8_t720}. At this stage the instantaneous eigenvalue distribution as well as FTLE distribution
is more comparable with the usual spectrum of edge state solutions, see Fig.~\ref{fig:insteigs_short2}: one dominant unstable eigenvalue marking a direction locally transversal to the edge manifold, several weakly positive eigenvalues expressing the chaotic nature of the edge state fluctuations, and (not appreciable here because of the small value of $r$) a large set of stable eigenvalues expressing the attraction of the edge state within the edge manifold.

One clear feature from physical space visualisations, regardless of the quantity plotted, is how the localised support of all OTD modes, except here for $u_8^{\lambda}$, superimposes exactly with the location of the edge state. This suggests that the present modes, if they contribute to an instability of the edge state, would \emph{not} make the main coherent edge state spread spatially, at least at the level of the linearised dynamics. As far as the unsteady dynamics restricted to the edge manifold is concerned, this suggests that shift sideways are excluded near $t\approx 280$ whereas they are likely to occur at $t\approx 720$. Such sideways shifts have been reported in most edge states of boundary layer flows \citep{khapko2013localized,khapko2016edge,beneitez2019edge}, ASBL \cite{khapko2016edge} as well as channel flow \citep{toh_itano_2003}. As in the present case, the shift phases are usually short and alternate with long shift-free phases. Another robust feature of all localised edge states concerns transition from the edge state to the turbulent state: the transition process consists of two consecutive steps: first a local intensification of the disturbances within the active core, followed by spatial spreading \cite{mellibovsky2009transition,duguet2010slug,Duguet2013}.

The consecutive nature of these two events would suggest that the spreading phase is nonlinear, while the intensification phase can be understood partially from the linear instability of the edge state. The fact that the spreading is reflected in the spatial structure of at least one instantaneous eigenmode $u_8^{\lambda}$ at the later time $t=$720, suggests however that spanwise spreading can be partially predicted and described at this time by linear mechanisms. These new results suggests further study.

\section{Conclusion and outlooks}

We have used the recently developed framework of the Optimally Time-Dependent (OTD) modes to study the linearised dynamics about a segment from a well-defined \emph{unsteady} base flow. The methodology was applied to a complex hydrodynamic case at the limit of our computational capabilities, and yields results in line with the expected physics. It even performs beyond expectations by revealing new physical phenomena. The physical system under investigation is the Blasius boundary layer flow.  The original trajectory under scrutiny belongs by construction to the edge manifold delimiting bypass from natural transition. However, the study is restricted to timespans short enough such that Tollmien-Schlichting waves do not have time to affect the transition process. This unsteady trajectory is re-interpreted as an unsteady base flow, whose linear (modal) stability analysis is expected to contain information about the stability of localised streaks, as observed in instances of bypass transition. This choice of base flow, due to its three-dimensionality and its unsteady dynamics, represents an excellent test case for a new stability approach.

 Limiting ourselves, as a computational compromise, to a projection basis consisting of only 8 OTD modes, we have computed the instantaneous eigenvalues along the unsteady trajectory. The streaky base flow displays a couple of unstable complex conjugate eigenvalues which dominate the finite-time stability of the trajectory. The remaining eigenvalues investigated have a positive real part as well, yet with a smaller magnitude. This is consistent with the expectations for chaotic dynamics within the edge manifold, although the notion of chaos is usually kept for the infinite-time frameworks.
 
 Numerical evidence suggests that the leading instability mechanism(s) in this study correspond to an outer mode as described by \cite{vaughan2011stability,hack2014streak}, even if the corresponding perturbations lack the long wavelength structure characteristic of streak eigenmodes reported so far \citep{andersson2001breakdown,brandt2014lift}. We have also analysed the Finite-Time Lyapunov exponents (FTLEs) along the trajectory by considering several time horizons. The results confirm the presence of one fast unstable direction versus many slower state space directions. Moreover, we could confirm that the underlying invariant set has a finite-time fractal dimension strictly larger than 8.

The leading modal structures obtained from the OTD modes
are not trivial to describe, mainly due to the lack of spatial symmetry of the base flow. The main property exploited in this study regards the spatial localisation of the modes. Most of the modes computed for $r$=8 display, in an instantaneous fashion, the same localisation properties as the original base flow. The most unstable perturbations display a positive instantaneous growth rate, and their vortical activity is classically located in the region adjacent to the streaks, where the total shear is highest \citep{schmid2001stability}. In particular, the perturbations in the $xy$-plane are tilted from the wall by an angle larger than the base flow, particularly at larger times. Some of the modal perturbations extracted also display vortical fluctuations upstream of the base flow, while one identified mode even displays localisation on the spanwise side of the base flow (at a later time only). It is suggested that the latter eigenmode plays an active role as precursor in streak-switching events, the same events that lead the localised edge state to propagate sideways. Downstream fluctuations are however absent from the leading instantaneous eigenmodes, suggesting that they are not fundamental to the temporal sustainment of the edge state \citep{Duguet2012}.

Although the method originally targets a modal description of the relevant finite-time instabilities, it can also capture non-normal amplification mechanisms \citep{babaee2016minimization}. In practice the exact amount of non-normality predicted, as well as the associated energy amplification, are constantly underestimated for finite $r$ compared to the full-dimensional problem, mainly because a larger number of instantaneous eigenmodes would be required to faithfully capture non-normal effects. Nevertheless these results confirm that non-normal effects also play a role in the streak breakdown phenomenon \citep{schoppa2002coherent,hoepffner2005transient}.

This study highlights the relatively large sensitivity, on short times, of the instantaneous eigenvalues to the initialisation of the OTD modes. It is expected from theoretical arguments \cite{babaee2017} that FTLEs can be safely computed from the eigenvalues only past a transient time, which is \emph{a priori} unknown and case-dependent. A detailed comparison between two different arbitrary initialisations suggests that, in the present case, only the early times prior to $t \approx$ 50 are highly dependent on the choice made for $t=0$ (cf.\ figure \ref{fig:IC_evals_check}. Although this transient can be considered as short relative to the complete transition process, it still represents a clear limitation of the method as far as  early times are concerned. At times larger than 50, the instantaneous eigenvalues $\lambda_1, \ldots$ evolve qualitatively similarly with time although instantaneously eigenvalues may differ between the two simulations. The corresponding trend is also valid for the numerical abscissa $\sigma$. If the dynamics belonged to an attractor, the time-averaged FTLEs would converge to the LEs, known to be independent of the initialisation \cite{pikovsky2016lyapunov}. Although the present case does not revolve around a genuine attractor in state space, the results in figure \ref{fig:IC_evals_check} 
clearly suggest that the late-time dynamics can be considered as temporally converged. Note that for $r$ large enough the discrepancy between different initialisations is expected to vanish even at finite times, for instantaneous eigenvalues as well as for the numerical abscissa. However, additional modes (and thus larger $r$) also imply a significant increase in computational time. 
 
On the technical level, several points require further discussion and study: 
\begin{itemize}
    \item (i) the size of the OTD subspace cannot be determined \emph{a priori}  \citep{babaee2016minimization,kern2021}. This is in particular relevant to capture the non-normality along the reference trajectory, where a large number of modes are required. Note that in cases with extensive systems, or ``weak turbulence'', such as Kuramoto-Sivashinsky \citep{cvitanovic2010state} just a few modes are required to entirely describe the most unstable subspace, whereas in pulsating Poiseuille flow more than 70 modes are required to fully describe the non-normal behaviour \citep{kern2021}. However, it has been shown that much lower number of modes $r\approx 6$ could already bring relevant physical insight \citep{kern2021}
 \item (ii)  eigenvalue crossing can make the OTD basis readapt multiple times
  \item (iii) the modal structures arising from the OTD framework are not associated with a single mode in the sense of classical linear stability analysis. The projected OTD modes contain information about several different mechanisms taking place at the same time, in particular for very complex reference trajectories. 
  \end{itemize}

To further clarify the potential of the OTD modes in the present complex flow case, we gather our main results in the following list :
\begin{itemize}
    \item first demonstration that the stability analysis of unsteady trajectories is technically possible for a complex large-scale system, without resorting to average Lyapunov exponents or Covariant Lyapunov vectors, even in the case where those might not be available.
    \item evidence that one unstable mode dominates over all the others at all times, a feature not at all obvious for an aperiodic flow.
    \item quantification of finite-time Lyapunov exponents along the edge trajectory, including the early times.
    \item quantification of the growth of state space volumes as time progresses, showing that at later times the requested number of modes is reduced compared to earlier times.
    \item first quantitative evidence for non-normal effects in an aperiodic flow.
    \item occurrence of spanwise shifts detected in the higher-order modes at late times.
    \item evidence that the sinuous symmetry prevails over the streamwise-independent structures throughout all the study.  In particular varicose perturbations, known as an alternative way to break streamwise independence and popularised by hairpin vortex studies, appear absent from our study.
    \item evidence that the new modes found in the present analysis can also be described as outer modes.
\end{itemize}

Looking ahead, although for intermediate times the OTD modes capture the non-normal features of the underlying linear dynamics, for large times the proposed methodology (edge tracking together with linear stability analysis using OTD modes) is essentially a generalisation of modal stability analysis to unsteady cases. Persistent consequences of the non-normality include for instance the finite-time instabilities likely to occur during the Orr mechanism (for $t<60$) and the lift-up at later times. Both require further extensions of this methodology for a quantitative prediction. The optimal framework proposed by \cite{schmid2007nonmodal} is intrinsically non-modal, and it is well suited to the identification of the disturbance most amplified in finite time over an unsteady base flow. The corresponding adjoint-looping algorithm was used successfully by \cite{cossu2007optimal} in channel flow, except that the reference trajectory chosen was not an edge trajectory but a linear transient. It would be interesting to apply the same methodology on an unsteady edge trajectory and compare the results with the present ones, to see whether one of the methods can predict the finite-time growth of coherent structures not captured by the other technique. Moreover, the possibility of combining these several techniques together is interesting for future developments in stability analysis.

\acknowledgements

M.B. and Y.D. would like to thank Hessam Babaee and Simon Kern for discussions about the OTD modes. Financial support by the Swedish Research
Council (VR) grant no. 2016-03541 is gratefully acknowledged. The computations were enabled by resources provided by the Swedish National Infrastructure for Computing (SNIC) partially funded by the Swedish Research Council through grant agreement no. 2018-05973.
The authors report no conflict of interest.

\appendix

\section{Further computational details}

\label{sec:implementation}

This appendix provides further details about the computation of the OTD modes using a pseudospectral approach, and in particular using the SIMSON code \citep{chevalier2007simson}. Consider the equations for the evolution of the OTD modes about a trajectory evolved with the Navier--Stokes equations: 
\begin{equation}
    \mathbf{\dot{u}}_i = \mathbf{L}_{\text{NS}} (\mathbf{u}_i)-\langle \mathbf{L}_{\text{NS}} (\mathbf{u}_i), \mathbf{u}_i \rangle \mathbf{u}_i-\sum_{j=1}^{i-1} [\langle \mathbf{L}_{\text{NS}} (\mathbf{u}_i), \mathbf{u}_j\rangle + \langle \mathbf{L}_{\text{NS}} (\mathbf{u}_j), \mathbf{u}_i\rangle ]\mathbf{u}_j, \label{eq:otdNS}
\end{equation}
where $\mathbf{L}_{\text{NS}}$ denotes the linearised Navier--Stokes operator. Bold letters denote quantities in physical space, which are discretised into $\mathbb{R}^{n}$ degrees of freedom. The linearised Navier--Stokes functional without external forcing applied to a field $\mathbf{u}_i$ reads,
\begin{equation}
    \mathbf{L}_{\text{NS}} (\mathbf{u}_i) = 
        - (\textbf{U}_b \cdot \nabla ) \textbf{u}_i - (\textbf{u}_i \cdot \nabla) \textbf{U}_b - \nabla p_i + \frac{1}{\mbox{Re}} \nabla^2 \textbf{u}_i. \label{eq:lns}
\end{equation}

The additional constraint to the linearised Navier--Stokes equations is introduced in SIMSON in the form of an explicit forcing at each time step. Boundary conditions are implemented into the linear part of the solver, while the nonlinear terms are evaluated explicitly. In the present case, evaluating explicitly the inner products involving the linearised Navier--Stokes operator can produce erroneous results if the boundary conditions on the additional forcing term are not applied properly. The term $\mathbf{L}_{\text{NS}}(\mathbf{u}_i)$ needs to contain the boundary conditions corresponding to the linearised operator to provide a correct forcing term and ${\bm L}_r$ in the computations. In particular, it is necessary to apply Neumann boundary conditions on the free stream and Dirichlet boundary conditions at the wall
\begin{equation}
    {\bm u}_i(y=0)=\frac{\partial {\bm u}_i}{\partial y}(y=L_y) = 0 
\end{equation}
when recovering the wall-normal velocity from the 4th order equation arising from the velocity-vorticity formulation \citep{chevalier2007simson}. This differs from the main body of the implementation in SIMSON since the correct boundary conditions need to be applied in the explicit term involving $\mathbf{L}_{\text{NS}}(\mathbf{u}_i)$ as well as the implicit part of the solver.

The OTD modes converge exponentially fast to the most unstable directions of the Cauchy-Green tensor \citep{babaee2017}, and after a long time only depend on the point of the trajectory where they are computed \citep{blanchard2019analytical}. However, the OTD modes depend on their initialization \citep{babaee2016minimization,kern2021}. It has been observed that there is not an universal time for which the OTD subspace is converged to the most unstable directions. Nevertheless, relevant physical features may be observed from early times \citep{babaee2016minimization}.

\section{Initial conditions for the OTD modes}
\label{sec:ICs_appendix}
An additional point to consider is that, if one of the directions  not part of the basis becomes unstable enough, the basis will need to re-adapt. This is due to eq. \eqref{eq:otd} being evolved continuously, whereas the introduction of a different vector in the most unstable subspace occurs discontinuously \citep{babaee2016minimization}.  

\begin{figure}
    \centering
        \includegraphics[width=0.8\textwidth]{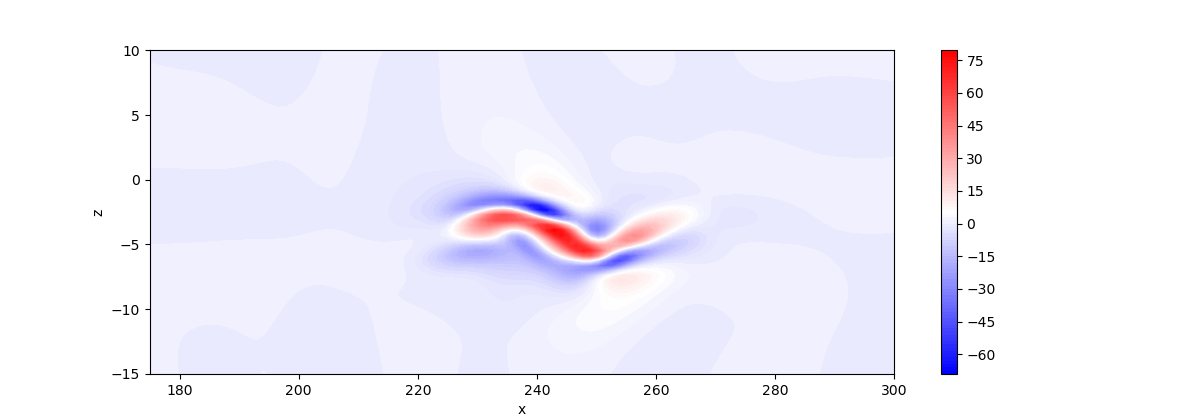} \\
        (a)\\
    \includegraphics[width=0.8\textwidth]{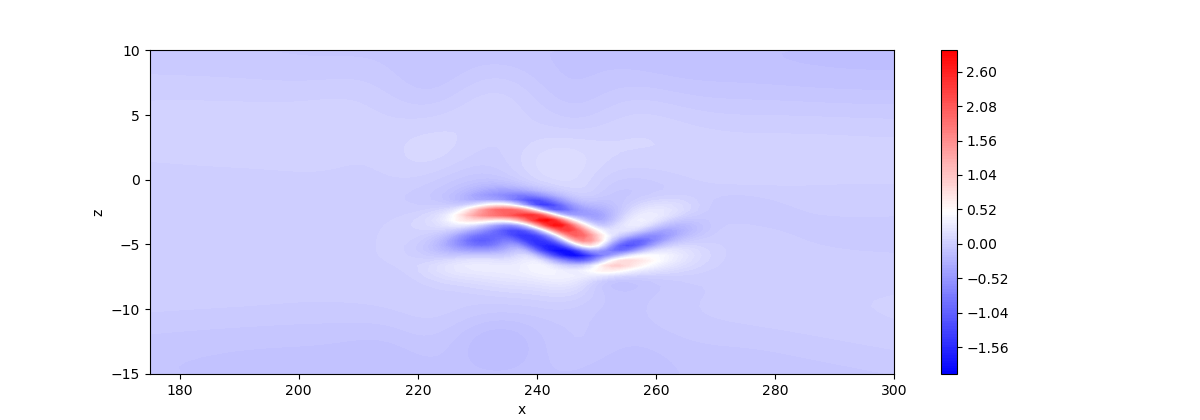}\\
    (b)
        \caption{$t=120$. Contours of the wall-normal velocity $V$ for $u^{\lambda}_1$ obtained from two different sets of initial conditions. (a) Subset initial conditions used in the current work. (b) Alternative initial conditions described in the Appendix \ref{sec:app_IC}. Note that according to \eqref{eq:Ulambda}, the $\{u^{\lambda}\}$ are not normalised.}
    \label{fig:IC_comp}
\end{figure}
The choice of the initial condition for the OTD modes plays therefore a crucial role for the OTD framework. Since our reference trajectory consists of several finite-time events of interest, our goal is to choose initial conditions which adapt as quickly as possible to the most unstable dynamics. We therefore chose initial conditions which are physically relevant to excite instability mechanisms on the edge trajectory. The initial condition for the first mode is exactly the  perturbation to the Blasius boundary layer associated with the edge state. This represents infinitesimal perturbations of the same shape as the edge trajectory. The initial conditions for modes 2-8 correspond to pairs of counter-rotating vortices with different spatial extensions. The counter-rotating vortices are also rotated  about the $y$ axis to remove any symmetric constraint. This set of initial conditions is not orthogonal by construction and therefore a Gram-Schmidt algorithm is performed before initialising the OTD computations. The results reported in the body of the paper correspond to these initial conditions.

\label{sec:app_IC}

To check the robustness of the results, alternative sets of initial conditions have been tested using $r=4$: (i) The 4 leading modes from the results in the main body of the paper and (ii) random noise. Using $r=4$ modes only appeared sufficient to illustrate the main aspects of the subsequent checks.

A comparison for the wall-normal component of the leading projected OTD mode at $t=120$, obtained using the two different sets of initial conditions can be seen in Fig.~\ref{fig:IC_comp}. The figure shows an agreement about the general physical features of the perturbation, i.e. the high-speed streak flanked by two low-speed streaks is present in both cases. However, no exact match is observed. The convergence to an unique set of OTD modes is expected to be exponentially fast \citep{babaee2016minimization,babaee2017,blanchard2019analytical}, but the explicit times are strongly case dependent. \\

\begin{figure}
    \centering
    \begin{tabular}{cc}
        \includegraphics[width=0.49\textwidth]{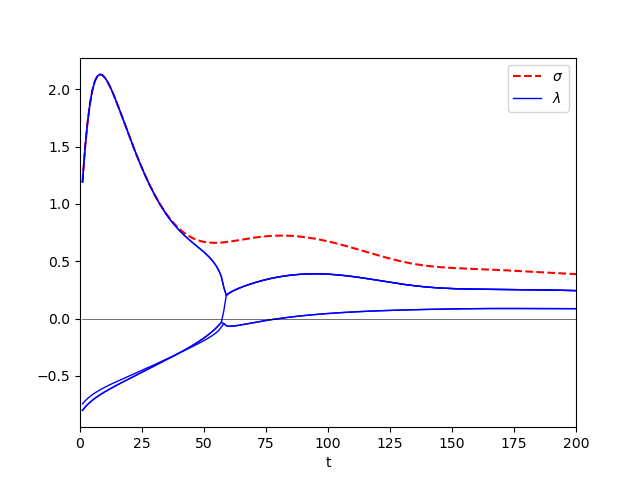} &
    \includegraphics[width=0.49\textwidth]{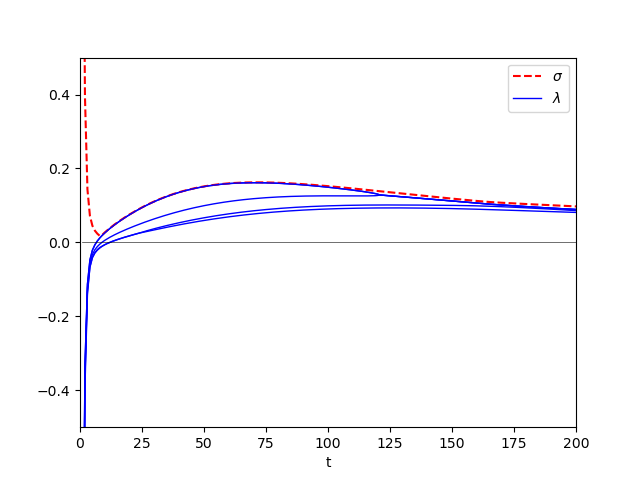}
\\
        (a) & (b) 
    \end{tabular}
        \caption{Instantaneous eigenvalues for $r=4$ corresponding to two different initialisations: (a) edge state perturbation over the Blasius boundary layer and counter rotating vortices, (b) random noise.}
    \label{fig:IC_evals_check}
\end{figure}

The most unstable instantaneous eigenvalues are shown in Fig. \ref{fig:IC_evals_check}. It can be observed that in the case of the random noise, the initial peak is lost. It is reasonable to assume that the unsteady base flow changes too fast during the initial times while the OTD subspace has not had enough time to adapt. On the other hand, the second peak identified at $t\sim 80$ is well identified with both sets of initial conditions. 

We should consider random noise as the worst choice of initial conditions, since it is entirely agnostic to the underlying reference trajectory. The results presented above further strengthen the importance of the choice of initial conditions. They indicates that, although the OTD approach is robust at large enough times, it remains dependent on the initialisation for times earlier than $t \approx 100$.

\bibliographystyle{jfm}
\bibliography{main}

\end{document}